\documentclass[letterpaper, 10 pt, conference]{ieeeconf}  

\IEEEoverridecommandlockouts                              

\usepackage{layout}
\usepackage{color}
\usepackage{xcolor}
\usepackage[]{graphicx}  
\graphicspath{{imgs/}}
\usepackage{booktabs}
\usepackage[caption=false, font=footnotesize]{subfig}
\usepackage{siunitx}
\usepackage{blindtext}
\usepackage{multirow}

\usepackage{url}
\usepackage{breakurl}
\usepackage[breaklinks]{}

\usepackage{makecell}
\usepackage{amsmath} 
\usepackage{amssymb}  
\usepackage[linesnumbered,ruled]{algorithm2e}

\title{\LARGE \bf
Polylidar - Polygons from Triangular Meshes*
}

\author{Jeremy Castagno$^{1}$ and Ella Atkins$^{2}$
\thanks{*This work was supported in part by NSF I/UCRC Award 1738714.}
\thanks{$^{1}$Jeremy Castagno is a Robotics Institute PhD candidate, University of Michigan, {\tt\small jdcasta@umich.edu}}%
\thanks{$^{2}$Ella Atkins is a Professor of Aerospace Engineering and Robotics, University of Michigan,
        {\tt\small ematkins@umich.edu}}%
}

\begin{document}

\maketitle
\thispagestyle{empty}
\pagestyle{empty}


\begin{abstract}
This paper presents Polylidar, an efficient algorithm to extract non-convex polygons from 2D point sets, including interior holes. Plane segmented point clouds can be input into Polylidar to extract their polygonal counterpart, thereby reducing map size and improving visualization. The algorithm begins by triangulating the point set and filtering triangles by user configurable parameters such as triangle edge length. Next, connected triangles are extracted into triangular mesh regions representing the shape of the point set. Finally each region is converted to a polygon through a novel boundary following method which accounts for holes. Real-world and synthetic benchmarks are presented to comparatively evaluate Polylidar speed and accuracy. Results show comparable accuracy and more than four times speedup compared to other concave polygon extraction methods.
\end{abstract}

\section{Introduction}

Video and LiDAR data are widely used in robotics to provide rich information about the environment. LiDAR and RGBD cameras generate point clouds for localization and mapping \cite{pathak_online_2010}, 3D modelling \cite{2016ISPAr49B3}, and scene classification for autonomous navigation \cite{Himmelsbach2010}. Flat surfaces such as walls and floors are key environmental elements to identify; they are often extracted using planar segmentation techniques \cite{feng_fast_2014, pham_geometrically_2016, schaefer_maximum_2019}.   However points clouds are dense incurring a high computational cost when used directly. A common simplifying approach transforms point clouds into lower dimensional representations such as lines and planes \cite{biswas_planar_2012}. Furthermore, polygonal representations of planes reduces map size and may accelerate matching for localization \cite{lee_indoor_2012}.
Convex polygon representations of planar segments were proposed by \cite{biswas_planar_2012}. Convex polygons are simple and efficient to generate but ignore boundary concavities and overestimate area of the enclosed point set. Non-convex polygon representations may be generated using techniques such as boundary following outlined in \cite{lee_indoor_2012} or $\alpha$-shapes as proposed in \cite{lee_fast_2013}. However few methods also capture the interior holes within non-convex polygons. Safe robot navigation demands accurate capture of non-convex polygons with interior holes in real-time, requiring both speed and robustness.



\begin{figure}[ht] 
    \centering
  \subfloat[]{%
    \centering
       \includegraphics[clip, trim=0.0cm 0.1cm 0.0cm 0.25cm, width=0.3\linewidth]{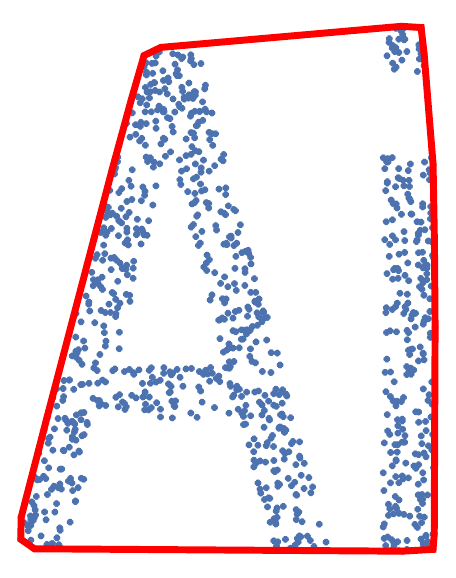}
    }
    \label{fig:convex}\hfill
  \subfloat[]{%
  \centering
        \includegraphics[clip, trim=0.0cm 0.1cm 0.0cm 0.25cm, width=.3\linewidth]{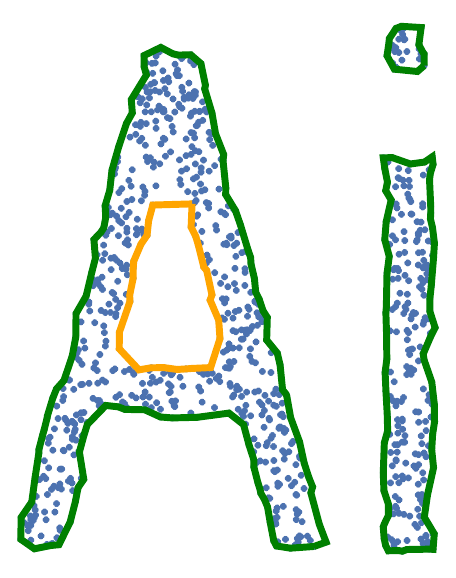}}
    \label{fig:concave} \hfill
  \subfloat[]{%
    \centering
      \includegraphics[width=0.35\linewidth]{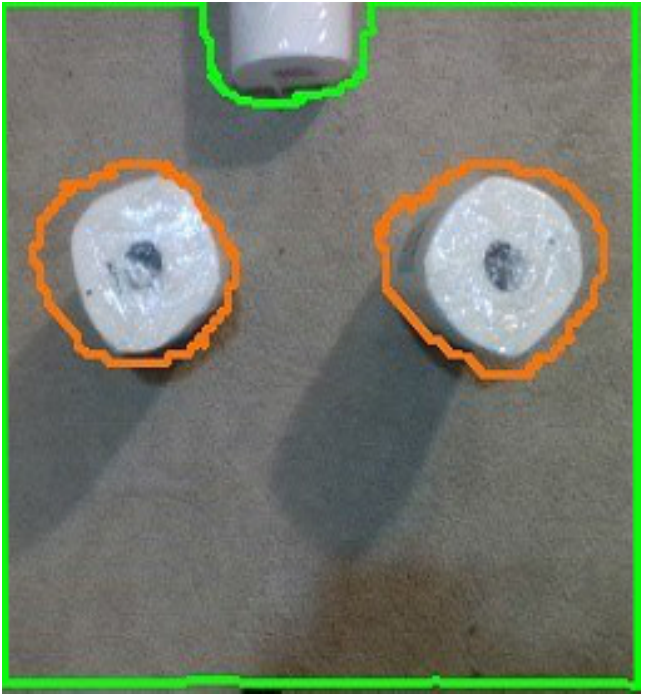}
    }
    \label{fig:realsense}\hfill\\
  \caption{(a) Convex hull of a point set (red); (b) MultiPolygon extraction using Polylidar (green).
  (c) Polygon extraction from a plane segmented point cloud from an Intel RealSense RGBD camera capturing paper towel rolls on a basement floor. 
  Note that Polylidar also identifies holes (orange).}
  \label{fig:convex_concave} 
\end{figure}

This paper presents Polylidar, an efficient algorithm to transform 2D point sets into simplified non-convex (i.e. concave) polygons with holes. Polylidar begins by triangulating the point set and filtering triangles given user-specified parameters such as maximum triangle edge length. Once filtering is complete, edge-connected triangles are combined into regions creating a set of triangular meshes representing the shape of the point set. Next, Polylidar converts each mesh region to a polygon through a novel boundary following method which accounts for holes. 
Figure \ref{fig:convex_concave}b shows Polylidar applied to a 2D point set while (c) shows Polylidar used on a plane segmented point cloud from an RGBD image.

We show that the Polylidar algorithm is approximately 4 times faster than leading open source approaches for concave polygon extraction. Polylidar's speed is attributed to rapidly identifying boundary edges (shell and holes) and then performing  boundary following to ensure a valid polygon is returned. 
Contributions of this paper are:

\begin{itemize}
  \item A faster open source \cite{polylidarcode} concave (multi)polygon extraction algorithm from 2D point sets.
  \item A benchmark comparison of leading concave polygon extraction techniques in terms of accuracy and speed.
\end{itemize}

Below, Sections \ref{sec:background} and \ref{sec:prelim} provide background on non-convex shape generation and mathematical preliminaries, respectively. Section \ref{sec:methods} describes Polylidar algorithms, while Section \ref{sec:results} shows benchmark test results of Polylidar versus other methods.  Section \ref{sec:random_polygons_test} describes test results. Sections  \ref{sec:discussion} and \ref{sec:conclusion} provide discussion and conclusions.

\section{Background}\label{sec:background}

Characterizing the shape of a set of 2D points $\mathcal{P}$ has been a long-term focus of computational geometry research. A convex hull is defined as the smallest convex polygon that fully encapsulates all points in a set $\mathcal{P}$.  Although widely used to estimate shape, point sets with non-convex distributions are poorly characterized by a convex hull \cite{Duckham2008}.  Convex hull over-estimation can be a serious issue when the points represent physical objects, e.g., obstacle free navigable areas. Several algorithms have been developed to construct shapes that ``fit'' or ``cover'' point sets more closely. 


Figure \ref{fig:convex_concave} compares convex and concave hulls. Figure \ref{fig:convex_concave}b is the multipolygon output of Polylidar described below.
While there is a unique convex hull, there is no true or unique concave hull.  Concave hull algorithm implementations can also have different output types.  Some return only an unordered set of edges while others return a single polygon.  Some algorithms return multiple disconnected polygons (multipolygon), and some can generate holes inside a polygon.

The $\alpha$-shape algorithm is an early strategy to generate a family of shapes ranging from a convex hull to a point set  \cite{1056714}. The parameter $\alpha$ dictates the radius of a closed disk used to prune/remove area in the convex hull. This disk is allowed to move freely shaving off the excess shape until it finds points. When disk radius is large, ideally infinite, the convex hull is produced; when disk radius is infinitesimally small only the points remain. A common implementation of $\alpha$-shape organizes points using Delaunay triangulation and filters triangles whose circumcircle radius is less than $\alpha$.  The final shape is represented by the remaining edges and triangles.  Note that the $\alpha$-shape method creates multiple non-intersecting shapes with the possibility of holes. 


The geospatial software library Spatialite \cite{spatialite}, an extension to SQLite \cite{chen2010use}, contains a concave hull extraction procedure. The algorithm again starts with Delaunay triangulation then analyzes the distribution of each triangle's edge length to determine mean $\mu_l$ and standard deviation $\sigma_l$. Any triangle with edge length greater than $C \cdot \sigma_l  + \mu_l$ is removed, where $C$ is a user-defined parameter. The final geometry returned is the union of all triangles computed with GEOS, a high performance open source geometry engine. The output may be a multipolygon (i.e., multiple disjoint polygons) with the possibility of holes inside each. 

PostGIS is a geospatial database of computational geometry routines such as the concave hull method in \cite{postgis}. This algorithm first calculates the convex hull and then shrinks the hull by adjusting vertex connections to closer points which ``cave in'' the hull.  This process recursively shrinks a boundary until a user-specified percent reduction in area from the convex hull is achieved. The resulting shape is a single polygon with the possibility of holes. 

\begin{table}[h]
\centering
\caption{Concave Hull Extraction Methods}
\label{table:compare_alg}
\begin{tabular}{ccc}
\hline
Algorithm                                                   & Output                                                           & Holes? \\ \hline
\begin{tabular}[c]{@{}c@{}}CGAL $\alpha$-shape\end{tabular}   & \begin{tabular}[c]{@{}c@{}}unordered\\ set of edges\end{tabular} & Yes    \\
Spatialite                                                 & (multi)polygon                                                   & Yes    \\
PostGIS                                                       & polygon                                                          & Yes    \\
Polylidar (new)                                                   & (multi)polygon                                                   & Yes    \\ \hline
\end{tabular}
\end{table}

Table \ref{table:compare_alg} provides a summary  of the concave hull algorithms discussed above. The Computational Geometry Algorithms Library (CGAL) is  used as the implementation of the $\alpha$-shape method \cite{cgal:eb-19a}. Note that the time complexity of all algorithm implementations, with the exception of PostGIS, is $\mathcal{O}(n\log{}n)$.  Our paper contributes a procedure to more rapidly compute (multi)polygon output with the possibility of holes.  Though this is a complex output to generate, we show through benchmarks that our algorithm and implementation outperforms other available approaches.

\section{Preliminaries}\label{sec:prelim}

A 2D \textit{point set} is an arbitrarily ordered set of two dimensional points in a Cartesian reference frame. Each point is defined by orthogonal bases $\hat{\mathbf{e}}_x$ and $\hat{\mathbf{e}}_y$  with
\begin{equation}
\label{eq:point}
    \vec{{p}_{i}}=x\,\hat{\mathbf{e}}_x+y\, \hat{\mathbf{e}}_y= [x,y]
\end{equation}
where $x,y$ are plane coordinates.

An $n$-point array $\mathcal{P} = \{ \vec{{p}_{1}}, \vec{{p}_{i}}, \ldots, \vec{{p}_{n}} \}$ contains points $\vec{{p}_{i}} \in \mathbb{R}^2$ indexed by $i$.  A triangular mesh $ \mathcal{T}$ is defined by
\begin{equation}
\label{eq:tri}
    \mathcal{T} = \{ t_1, t_i, \ldots, t_{k} \}
\end{equation}
where each $t_i$ is a triangle with vertices defined by three point indices $\{i_1, i_2, i_3\} \in \left[1,n\right]$ referencing points in $\mathcal{P}$.

We follow the Open Geospatial Consortium (OGC) standard \cite{herring2006opengis} for defining \textit{linear ring} and \textit{polygon}. A linear ring is a consecutive list of points that is both closed and simple. This requires a linear ring to have non-intersecting line segments that join to form a closed path. The key components of a valid polygon are a single exterior linear ring representing the \emph{shell} of the polygon and a set of linear rings (possibly empty) representing \emph{holes} inside the polygon. 


\section{Methods}\label{sec:methods}

Sections \ref{sec:tri}, \ref{sec:triangle_filtering_2d}, and \ref{sec:mesh_extraction} describe the triangulation data structures, filtering, and mesh extraction respectively.  Section \ref{sec:polygon_extraction} describes polygon extraction. 
\subsection{Triangulation with Half-Edge Decomposition} \label{sec:tri}

\begin{figure}[!ht] 
    \centering
  \subfloat[]{%
       \includegraphics[width=0.41\linewidth]{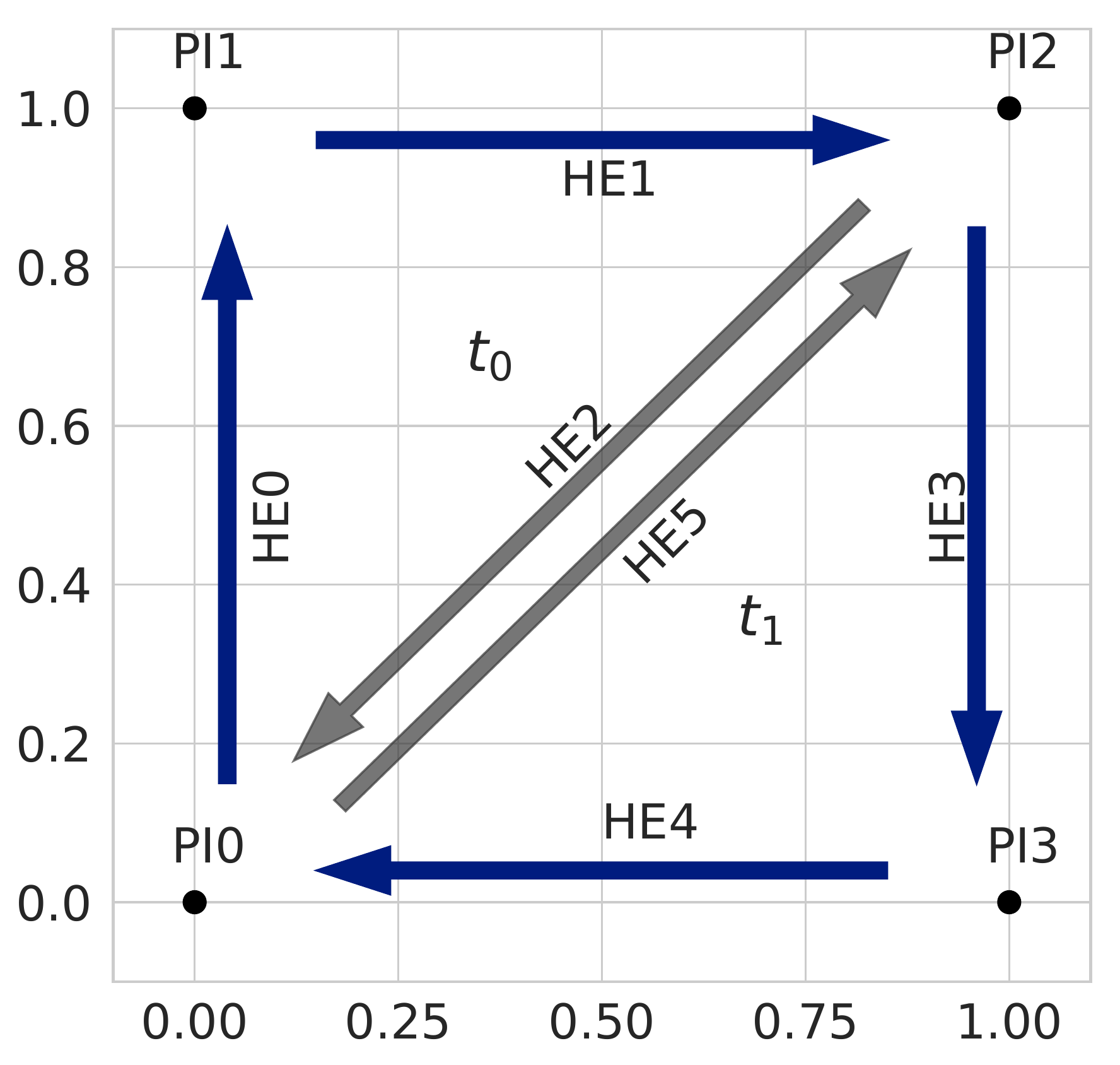}}
    \label{fig:delaunator_tri}\hfill
  \subfloat[]{%
        \includegraphics[width=.48\linewidth]{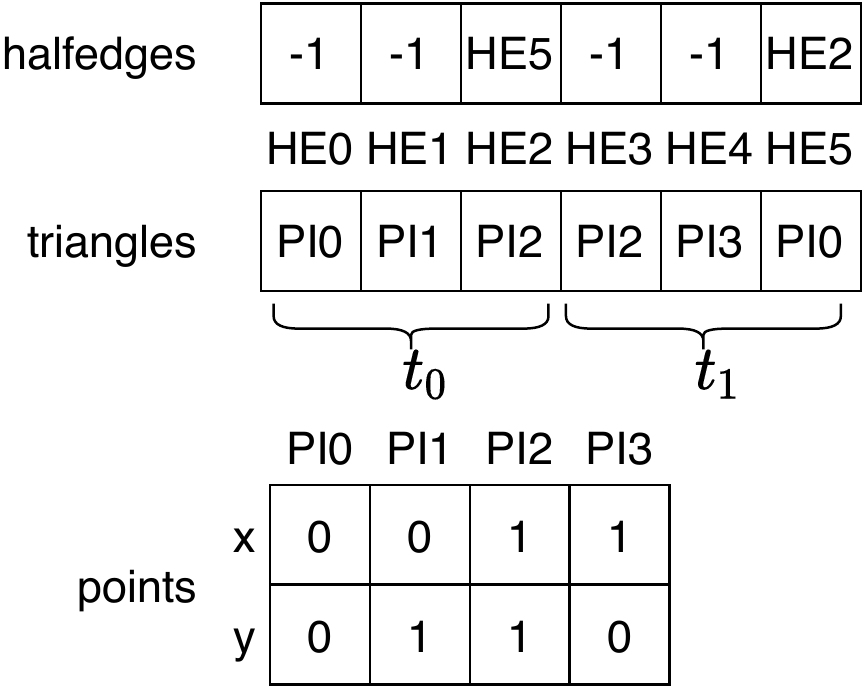}}
    \label{fig:delaunator_ds}\\
  \caption{(a) Triangulation of a square point set using Delauantor \cite{dealaunator} with output data structure indexed by half-edge ids in (b).  HE=half-edge, PI=point index, $t$=triangle. Grey edges show shared edges decomposed individually.}
  \label{fig:delaunator} 
\end{figure}

Polylidar begins with the Delaunator library \cite{dealaunator} performing a Delaunay triangulation of point set $\mathcal{P}$. The original algorithm was written in JavaScript but a C++ port of the library is used in Polylidar \cite{dealaunator_cpp}. Note that we have modified Delaunator to use robust geometric predicates to ensure correctness during triangulation \cite{shewchuk1997adaptive}. Delaunator was chosen for its ease of integration, speed, and output data structure which returns a \emph{half-edge} triangulation. A half-edge triangulation decomposes a shared edge using two half-edges A$\rightarrow$B and B$\rightarrow$A. An example of this decomposition and resulting data structures is shown in Figure \ref{fig:delaunator}.


Figure \ref{fig:delaunator}a triangulates point set $\{\text{PI0, PI1, PI2, PI3}\}$.  Triangulation produces two triangles, $t_0$ and $t_1$, with half-edges $\{\text{HE0, HE1, HE2}\}$ and $\{\text{HE3, HE4, HE5} \}$, respectively.  Each half-edge supports clockwise travel to the next half-edge in that triangle's edge set. Figure \ref{fig:delaunator}b lists the resulting $halfedges$, $triangles$, and $points$ data structures.  The $halfedges$ array is indexed by a half-edge reference id. It provides the opposite half-edge of a shared edge if it exists; otherwise -1 is returned. The $triangles$ array is also indexed by half-edge id and gives the starting point index of the associated half edge. The relationship between half-edge and triangle indices is $t = \operatorname{floor}(he / 3)$.

\subsection{Triangle Filtering}\label{sec:triangle_filtering_2d}
As with Spatialite and $\alpha$-shape methods the initial shape starts with $k$ triangles in $\mathcal{T}$ per Eqn. \ref{eq:tri} returned from Delaunay triangulation. Also similar to $\alpha$-shape and Spatialite methods, Polylidar filters triangles by configurable criteria for each triangle. Polylidar allows the user to perform triangle filtering using either the $\alpha$ parameter or maximum triangle edge length parameter $l_{max}$. The filtered triangle set is denoted $\mathcal{T}_f$. 

\subsection{Triangular Mesh Region Extraction}\label{sec:mesh_extraction}

An iterative plane extraction procedure inspired from \cite{doi:10.1080/01431161.2017.1302112} generates subsets of $\mathcal{T}_f$ that are spatially connected.  These subsets are denoted  $\mathcal{T}_r$ which represent triangular mesh \emph{regions}. A spatial connection between triangles exists when they share an edge. A random seed triangle is selected from $\mathcal{T}_f$ where a new region is created and expanded by its adjacent edge neighbors from the \emph{halfedges} data structure. Region growth halts when no more triangles in $\mathcal{T}_f$ connect to the region. The process repeats with another seed triangle until all triangles in $\mathcal{T}_f$ have been examined. 

\begin{figure}[!ht] 
    \centering
  \subfloat[]{%
       \includegraphics[width=0.40\linewidth]{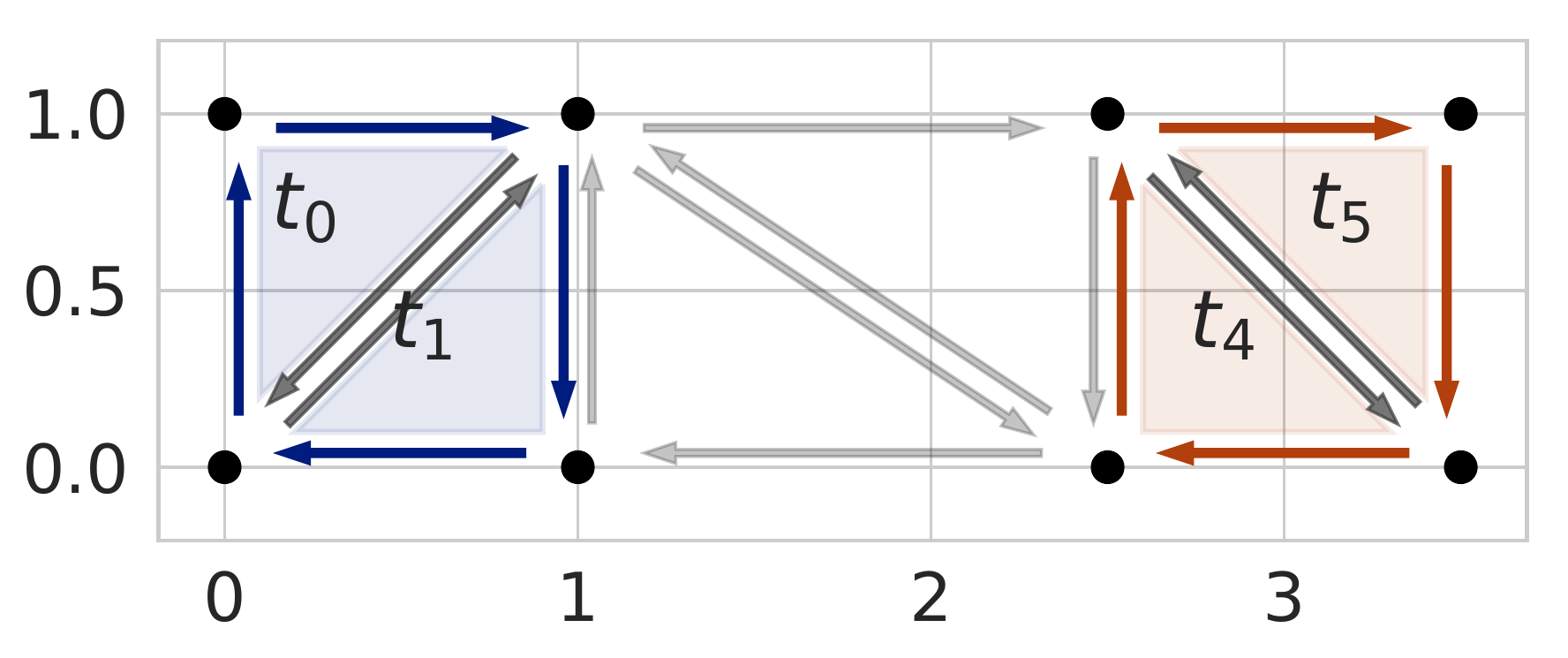}}
    \label{fig:plane_extraction_a}\hfill
  \subfloat[]{%
        \includegraphics[width=.49\linewidth]{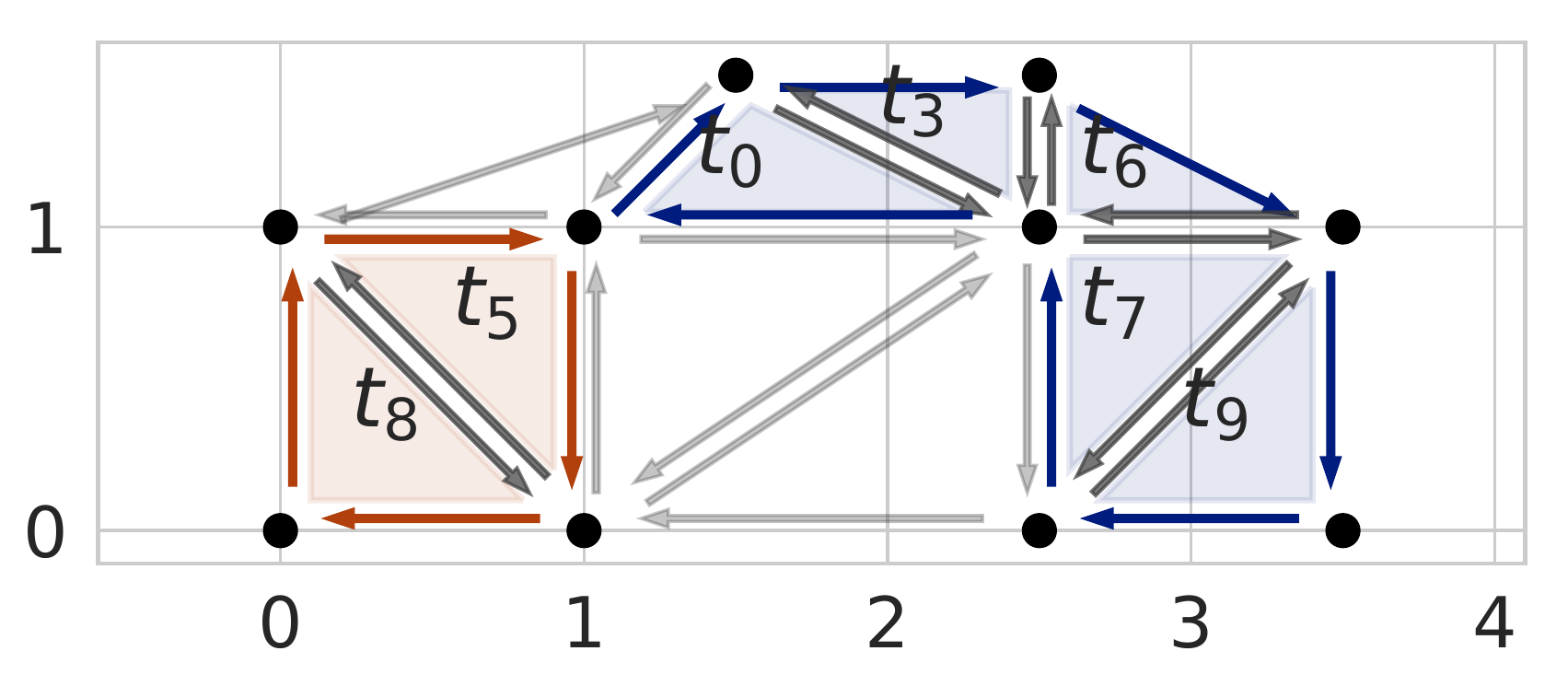}}
    \label{fig:plane_extraction_b}\\
  \caption{(a) Example of two regions extracted denoted by orange and blue. Triangles $t_0$ and $t_1$ are one region while $t_4$ and $t_5$ are another. (b) Two regions are also extracted even with a shared vertex.}
  \label{fig:plane_extraction} 
\end{figure}

 Figure \ref{fig:plane_extraction}a shows triangular mesh region examples. Distinct regions are shown in orange and blue; light grey edges denote triangles that have been filtered out.  
 The output of this step is a set of spatially connected triangular mesh regions, $\mathcal{T}_R$, where each specific region, $\mathcal{T}_{r,i}$, is a set of triangle \emph{indices}. We denote the set of $m$ triangular mesh regions as:

\begin{align}
    \mathcal{T}_R = \{ \mathcal{T}_{r,1}, \mathcal{T}_{r,i},  \ldots, \mathcal{T}_{r,m} \} \\
    \mathcal{T}_{r,i} =  \{ t_{i}, \ldots, t_{j} \}
\end{align}

\subsection{2D Polygon Extraction}\label{sec:polygon_extraction}

Polygon extraction has three steps: data structure initialization, concave shell extraction, and hole(s) extraction. Each of these steps is described below. Note that polygon extraction is independent of the specific triangular mesh regions $\mathcal{T}_{r,i}$, thus subsequent notation will drop the $i$ index for brevity when used in algorithms. The following steps are executed for each of the $m$ regions in $\mathcal{T}_{R}$ to generate $m$ polygons.

\subsubsection{Data structure initialization}

\begin{figure}[!ht] 
    \centering
  \subfloat[]{%
       \includegraphics[width=0.60\linewidth]{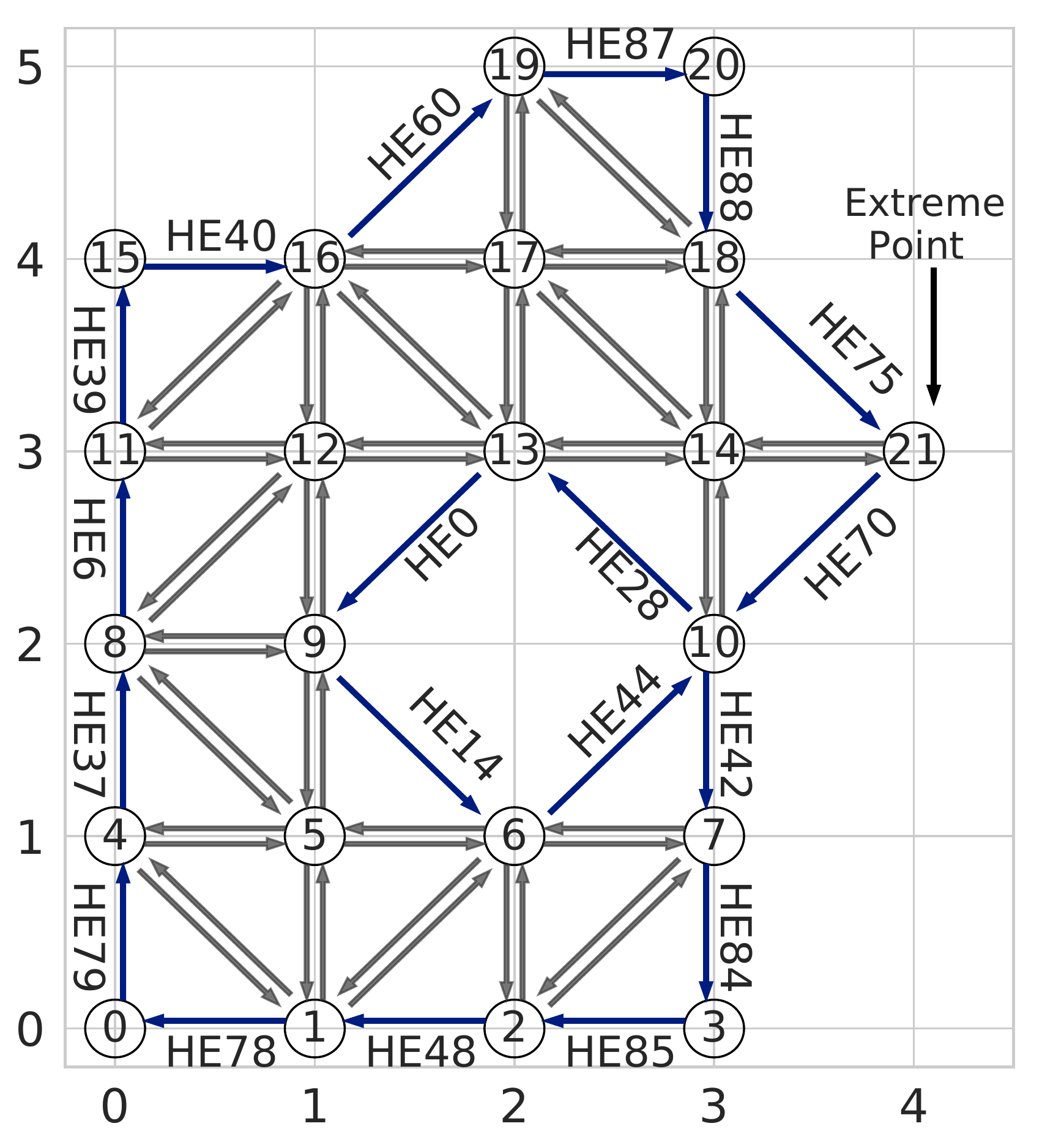}}
    \label{fig:algorithm1_visual_boundary}\hfill
  \subfloat[]{%
  \centering
        \includegraphics[clip, trim=0cm 0cm 0.5cm 0cm,width=.40\linewidth]{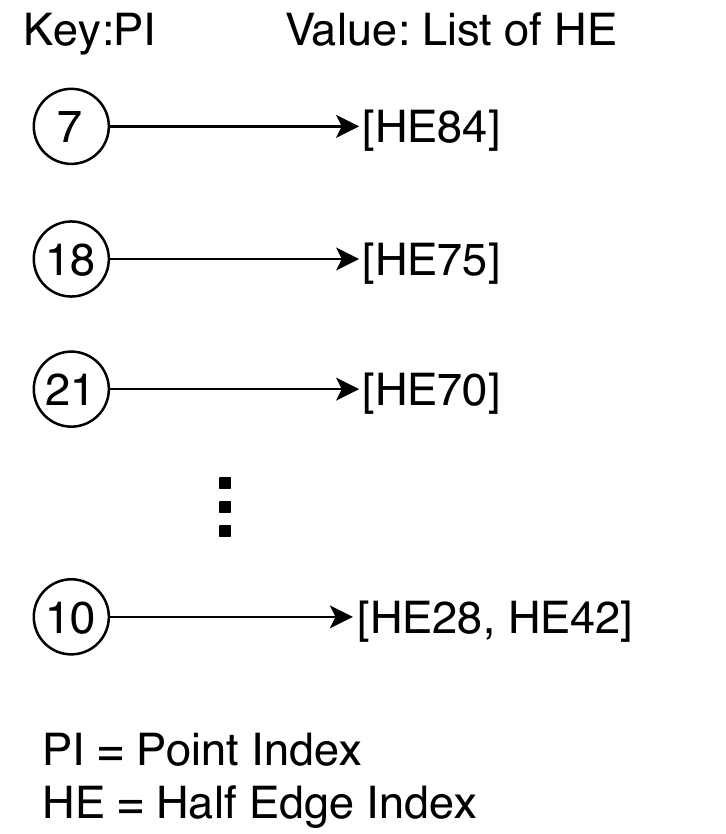}}
    \label{fig:hashmap}\\
  \caption{(a) The boundary half-edge set is marked in blue and point index 21 (PI21), the farthest point on the x-axis, is noted. (b) A sample of the resulting point index hash map, $PtE$ is shown. Note that the display order has been arbitrarily chosen.  }
  \label{fig:algorithm1_visual} 
\end{figure}

Data structure initialization is shown in Algorithm \ref{alg:boundary_edges} which produces three data structures: a boundary half-edge set, a point index hash map, and the extreme point. A visual example of these data structures is shown in Figure \ref{fig:algorithm1_visual}. Boundary half-edge set $\mathcal{HE}$ contains the half-edge indices that are on the exterior border of a region, marked in blue in Figure \ref{fig:algorithm1_visual}a.  A half-edge is marked as a boundary if it has no opposite half-edge (meaning it is on the convex hull of the full triangulated set) or if its adjacent triangle is not in $\mathcal{T}_{r,i}$. The last check is important because a half-edge may share an edge with an interior triangle that is not part of $\mathcal{T}_{r,i}$ as seen in the rightmost edge for the blue region in Figure \ref{fig:plane_extraction}a. The \emph{halfedges} data structure is fixed at triangulation and is not aware of filtered triangles or the regions discussed in Section \ref{sec:mesh_extraction}.

The second data structure is a point index hash map, $PtE$, whose \emph{key} is a point index and \emph{value} is a \emph{list} of outgoing boundary half-edges from the keyed point index. This \emph{unordered} hash map is represented in Figure \ref{fig:algorithm1_visual}b; note the keyed point index 7 mapping to the single element list containing half-edge 84. The final data structure represents an extreme point in the triangle mesh, referring to the point farthest to the right on the $x$-axis. This point will be used as the starting point index when extracting the concave hull to help ensure extraction does not start on a hole edge. Multiple points may exist on the extreme edge;  the algorithm will track the first one found in this case. 

\begin{algorithm}[t]\label{alg:boundary_edges}
    \SetKwInOut{Input}{Input}
    \SetKwInOut{Output}{Output}

    \Input{Triangular Mesh Region , $\mathcal{T}_r = \{t_i, \ldots, t_k\}$ \\
        Shared Halfedges, $halfedges$ \\
        Triangles Point Index, $triangles$ 
    }
    
    \Output{Half Edge Set , $\mathcal{HE} = \{he_i, \ldots, he_n\}$ \\ 
        Point Index Hash Map, $PtE$ \\
        Extreme Point, $pi_{xp}$ } 
    $\mathcal{HE} = \emptyset$ \tcp*{boundary half-edge set}
    $PtE = \emptyset$ \tcp*{Point to half-edge hashmap}
    $pi_{xp} = 0$ \tcp*{will be overwritten}
    \For{$t_i \in \mathcal{T}_r$} { 
        \For{$he_i \in t_i$} { 
            $he_j = halfedges[he_i]$\tcp*{opposite edge}
            $t_j = \operatorname{floor}(he_j / 3)$ \tcp*{adjacent tri}
            \uIf{$t_j \notin \mathcal{T}_r$}{
                $\mathcal{HE}$ = $\mathcal{HE}$ + $he_i$ \tcp*{boundary edge} 
                $pi = triangles[he_i]$ \\
                $pi_{xp} = \operatorname{TrackXp}(pi, pi_{xp})$ \\
                \uIf{$pi \notin PtE$}{
                    \tcc{create half-edge list}
                    $PtE[pi] = [he_i]$ 
                }
                \uElse {
                    $\operatorname{Append}(PtE[pi], he_i)$
                }
            }
        }
  
    }
    return $\mathcal{HE}, PtE, pi_{xp}$
    \caption{Initialize}
\end{algorithm}

\subsubsection{Concave Shell Extraction}
Outer shell extraction begins by traversing the half-edge graph, starting with the half-edge provided by the extreme point. As the edges are traversed the point indices are recorded in a list representing the linear ring of the concave hull. Edges are removed from the half-edge set, $\mathcal{HE}$, as they are traversed.  In Figure \ref{fig:algorithm1_visual}a the extreme point index is PI21 and the starting half-edge is HE70. This starting half edge and start point index are arguments to the \texttt{ExtractLinearRing} procedure in Algorithm \ref{alg:concave_hull_shell}, with the procedure halting when edge traversal returns back to the starting point index, indicating a closed linear ring has been extracted. The hole in this shape, represented by edges (HE28, HE0, HE14, HE44), with a shared vertex at PI10, must be carefully handled as explained below. This is an example of an non-manifold mesh.

\begin{figure}[ht] 
    \centering
  \subfloat[]{%
      \includegraphics[width=0.49\linewidth]{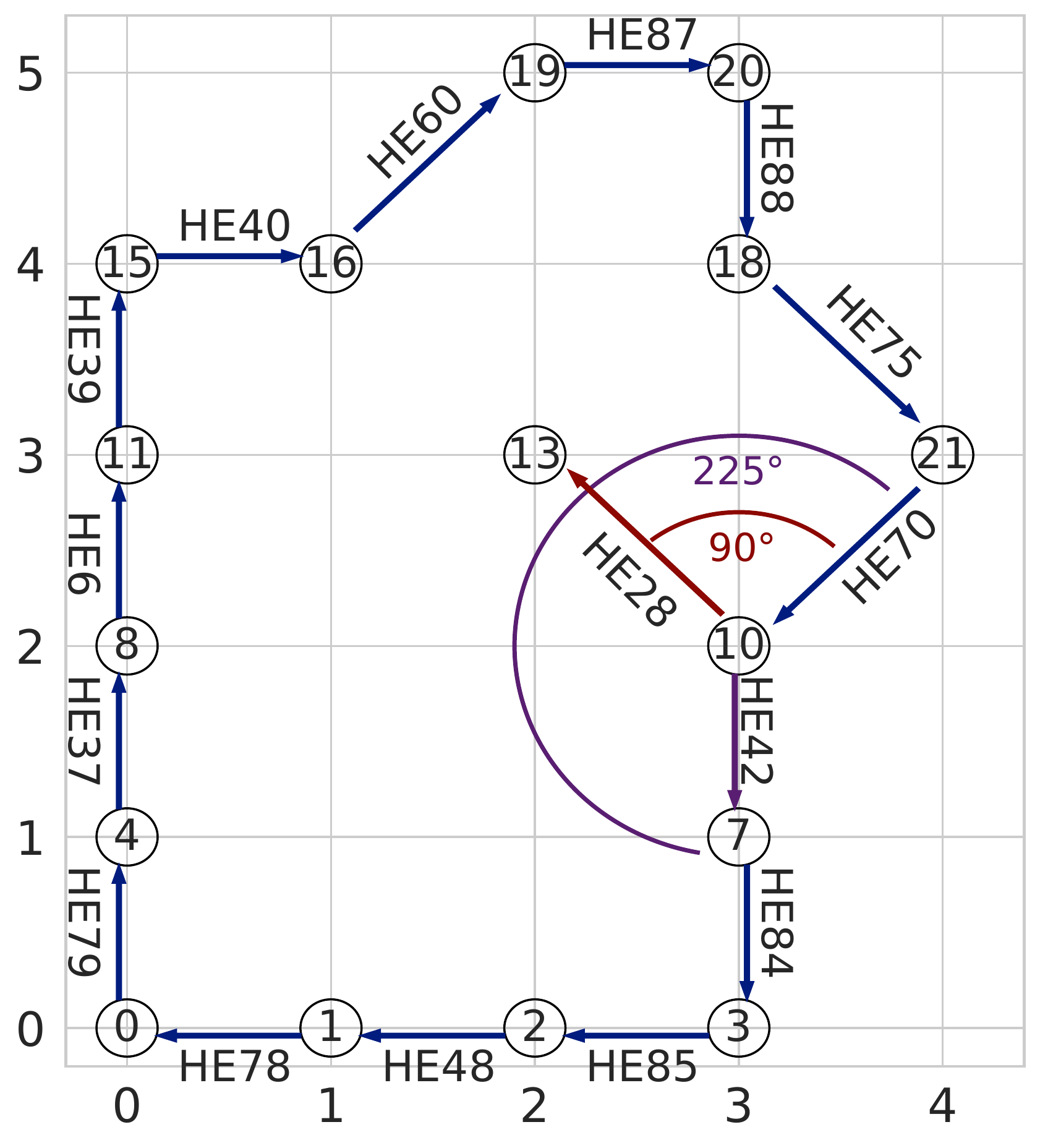}}
    \label{fig:example1_non_extreme} \hfill
  \subfloat[]{%
  \centering
        \includegraphics[width=.45\linewidth]{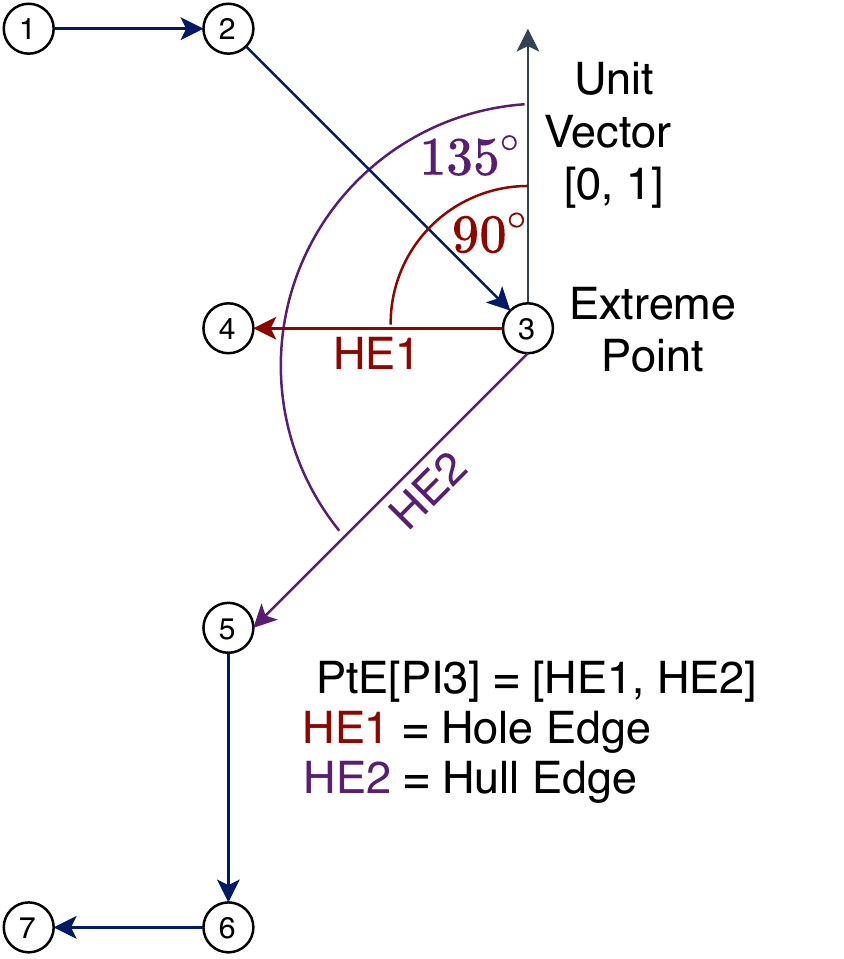}}
        \label{fig:example1_hole_extreme}
        \\
  \caption{(a) Edge selection for Fig. \ref{fig:algorithm1_visual}a.  HE70 leads to point index PI10 during shell extraction. Half-edges HE28 and HE42 leave PI10. The correct edge to follow (HE42) has the greatest angle with HE70. (b) If the extreme point has two outgoing edges (HE1, HE2), choose the edge with largest angle difference with the unit vector [0,1]. This is edge HE2.}
  \label{fig:example1} 
\end{figure}

The example in Figure \ref{fig:algorithm1_visual} begins with HE70 traversing to PI10. The outgoing boundary half-edges for this point index are determined from $PtE$ which provides a list of both HE28 and HE42. However HE28 is an edge for a hole in this polygon while HE42 is the correct half-edge to traverse for the outer shell. The \texttt{SelectEdge} procedure determines which of these edges to choose and is visually outlined in Figure \ref{fig:example1}a. Angles between the proposed edges and previous edge HE70 are calculated and the edge with the largest angle is chosen which guarantees the largest concave hull.  This edge cannot be a hole edge because that would imply that the hole is outside the concave shell, which is invalid.

On rare occasions the extreme point may have more than one outgoing half edge, meaning that a hole is connected to it.  This can be handled in the same way stated above by using the \texttt{SelectEdge} procedure. The only difference is that the previous hull edge is not known (the procedure has just started), but since we know we are on the far right of the hull we can substitute the previous edge for the unit vector [0,1] per Figure \ref{fig:example1}b. This unit vector is guaranteed to provide a stable order of the angle differences which would have been provided by the actual previous hull edge.

\begin{algorithm}\label{alg:concave_hull_shell}
    \SetKwInOut{Input}{Input}
    \SetKwInOut{Output}{Output}

    \Input{Half Edge Set , $\mathcal{HE} = \{he_i, \ldots, he_n\}$ \\ 
            Point Index Hash Map, $PtE$ \\
            Starting half-edge, $he$ \\
            Start point index, $startPI$ \\
            Triangles Point Index, $triangles$
    }
    \Output{Linear Ring , $lr = [pi_1, \ldots, pi_k]$ } 
    $lr = [\;]$  \tcc*{empty linear ring}
    \While{True}{
        $\mathcal{HE} = \mathcal{HE} \setminus he $ \\
        $he_t = \operatorname{NextTriangleEdge}(he)$\\
        $pi = triangles[he_t]$ \\
        $\operatorname{Append}(lr, pi)$\\
        \uIf{$pi \; is \; startPI$}{
            \tcc{closed linear ring}
            break
        }
        $nextEdges = PtE[pi]$ \\
        $he = \operatorname{SelectEdge}(he, nextEdges)$
    }
    return $lr$
    \caption{ExtractLinearRing}
\end{algorithm}

\subsubsection{Hole(s) Extraction}

\begin{figure}[ht] 
    \centering
  \subfloat[]{%
      \includegraphics[width=0.45\linewidth]{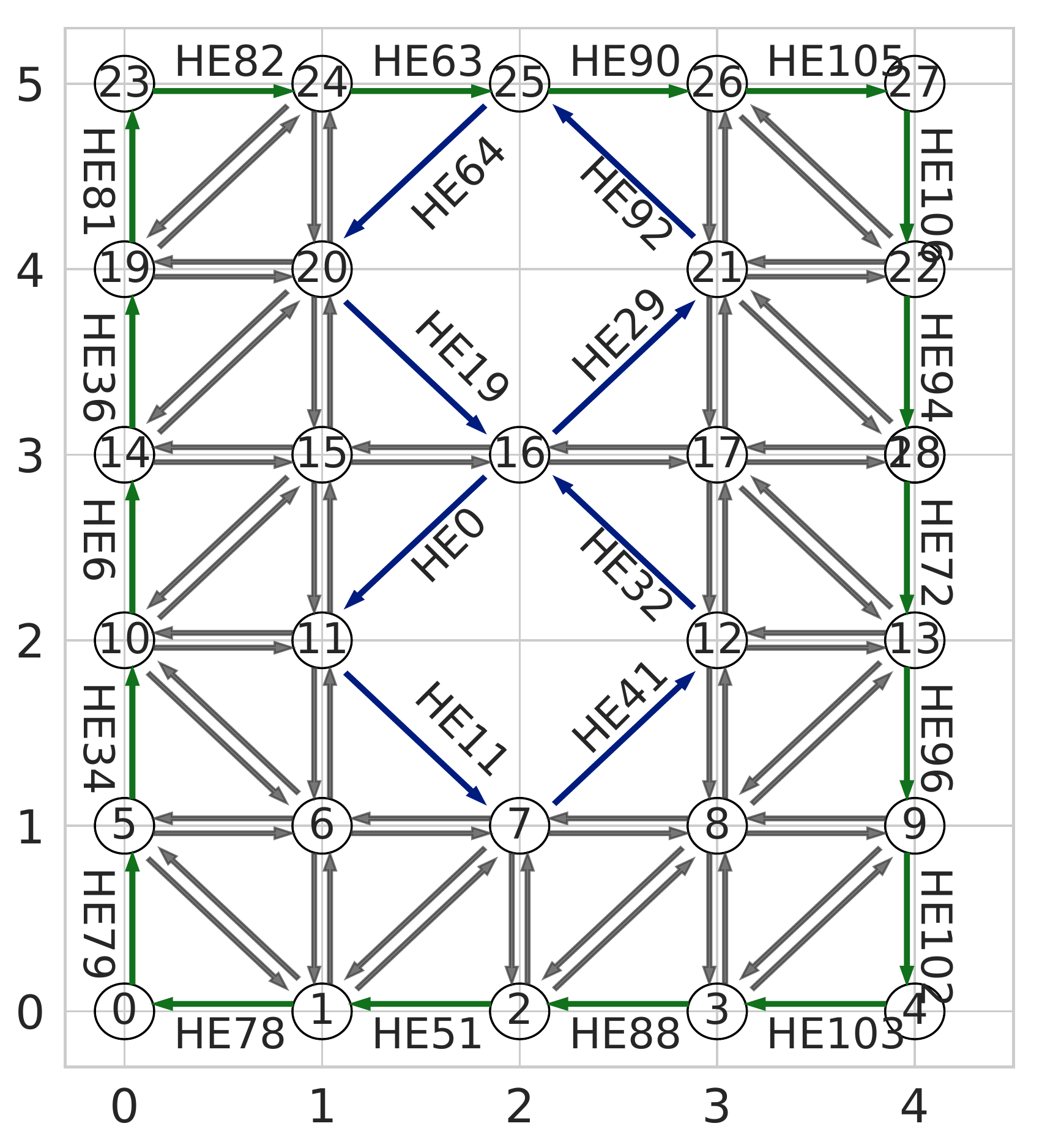}}
    \label{fig:example2_hull}\hfill
  \subfloat[]{%
  \centering
        \includegraphics[width=.45\linewidth]{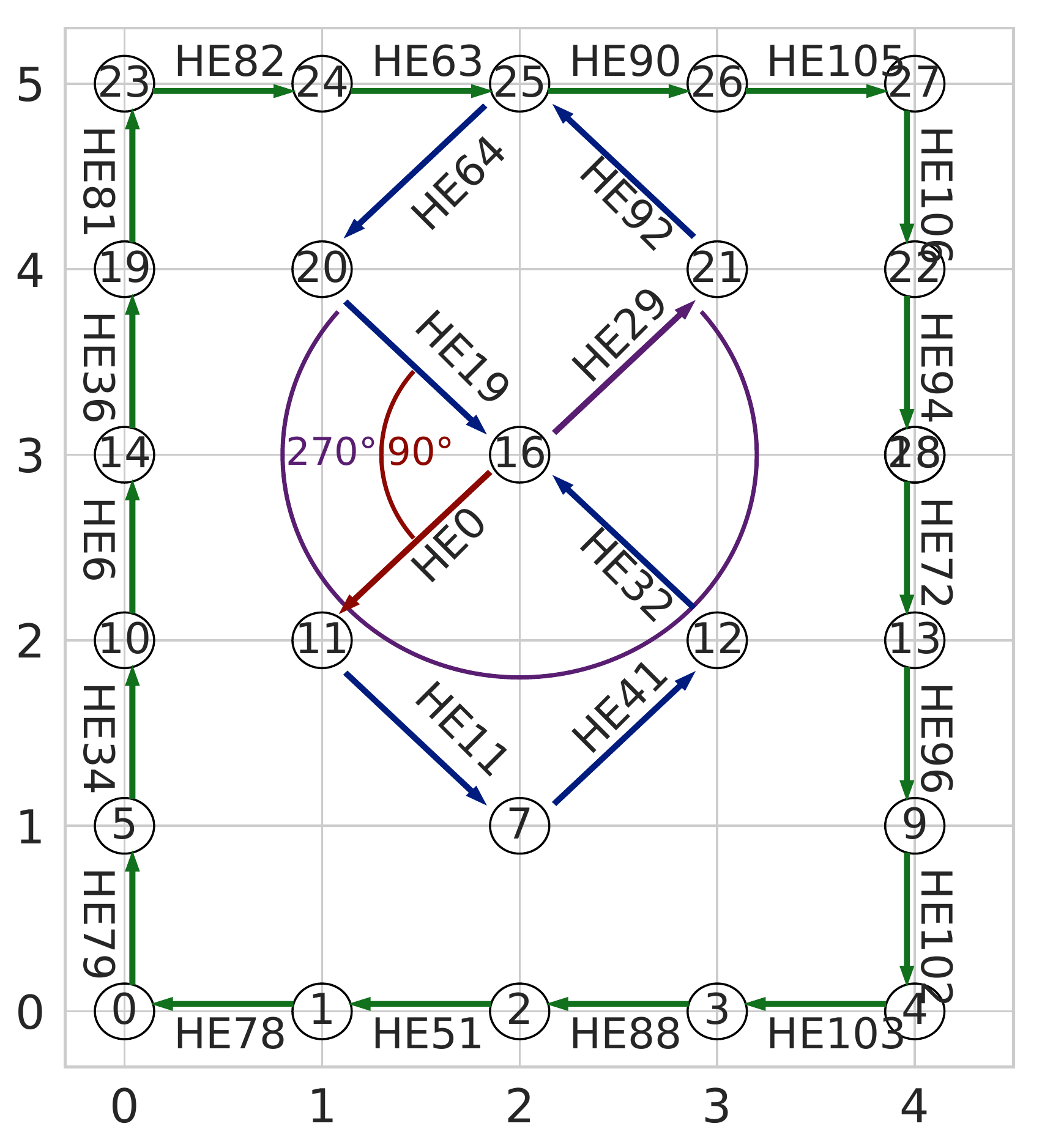}}
    \label{fig:example2_choice}\\
  \caption{(a) Edge case of two holes sharing the same vertex at PI16. The outer shell (green) is already extracted. (b) When traversing from HE19 to point index PI16, two outgoing edges (HE0 and HE29) are found. Edge HE29  with the largest angle difference from HE19 is chosen. }
  \label{fig:example2} 
\end{figure}

After the outer shell of the concave hull has been determined, only the holes remain to be found (if any holes exist). Any edges that remains inside $\mathcal{HE}$ are hole edges and will be extracted using Algorithm \ref{alg:hole_extraction}.  A half-edge is randomly chosen from $\mathcal{HE}$ for which the same \texttt{ExtractLinearRing} procedure is run. Figure \ref{fig:example2}a shows a corner case of a non-manifold mesh that must be handled if two holes share the same vertex. The previously extracted concave shell is displayed in green while the remaining half-edges to be processed are in blue; note the shared vertex at PI16.  Figure \ref{fig:example2}b shows the event when HE19 is randomly chosen for hole extraction leading to PI16. HE0 or HE29 is chosen in the manner previously discussed: the edge with largest angle guarantees the smallest hole thus is chosen.  If the other edge was chosen this would indicate a hole inside a hole which is invalid.

\begin{algorithm}\label{alg:hole_extraction}
    \SetKwInOut{Input}{Input}
    \SetKwInOut{Output}{Output}

    \Input{Half Edge Set , $\mathcal{HE} = \{he_i, \ldots, he_n\}$ \\ 
            Point Index Hash Map, $PtE$ \\
            Triangles Point Index, $triangles$
    }
    \Output{Set of Linear Ring Holes , $\mathcal{HR} = \{lr_1, \ldots, lr_k\}$ } 
    $\mathcal{HR} = \emptyset$  \tcc*{empty hole set}

    \While{$\mathcal{HE}$ is not empty}{

        $he= \operatorname{RandomChoice}(\mathcal{HE})$ \\
        $pi = triangles[he]$\\
        $lr = \operatorname{ExtractLinearRing}(\mathcal{HE}, PtE, he, pi, triangles)$ \\
    
        $\mathcal{HR} = \mathcal{HR} + lr$ 
    }
    return $\mathcal{HR}$
    \caption{Extract Holes}
\end{algorithm}

\begin{figure*}[!ht] 
    \centering
  \subfloat[]{%
      \includegraphics[clip, trim=1.5cm 0cm 1.5cm 0cm, width=0.28\linewidth]{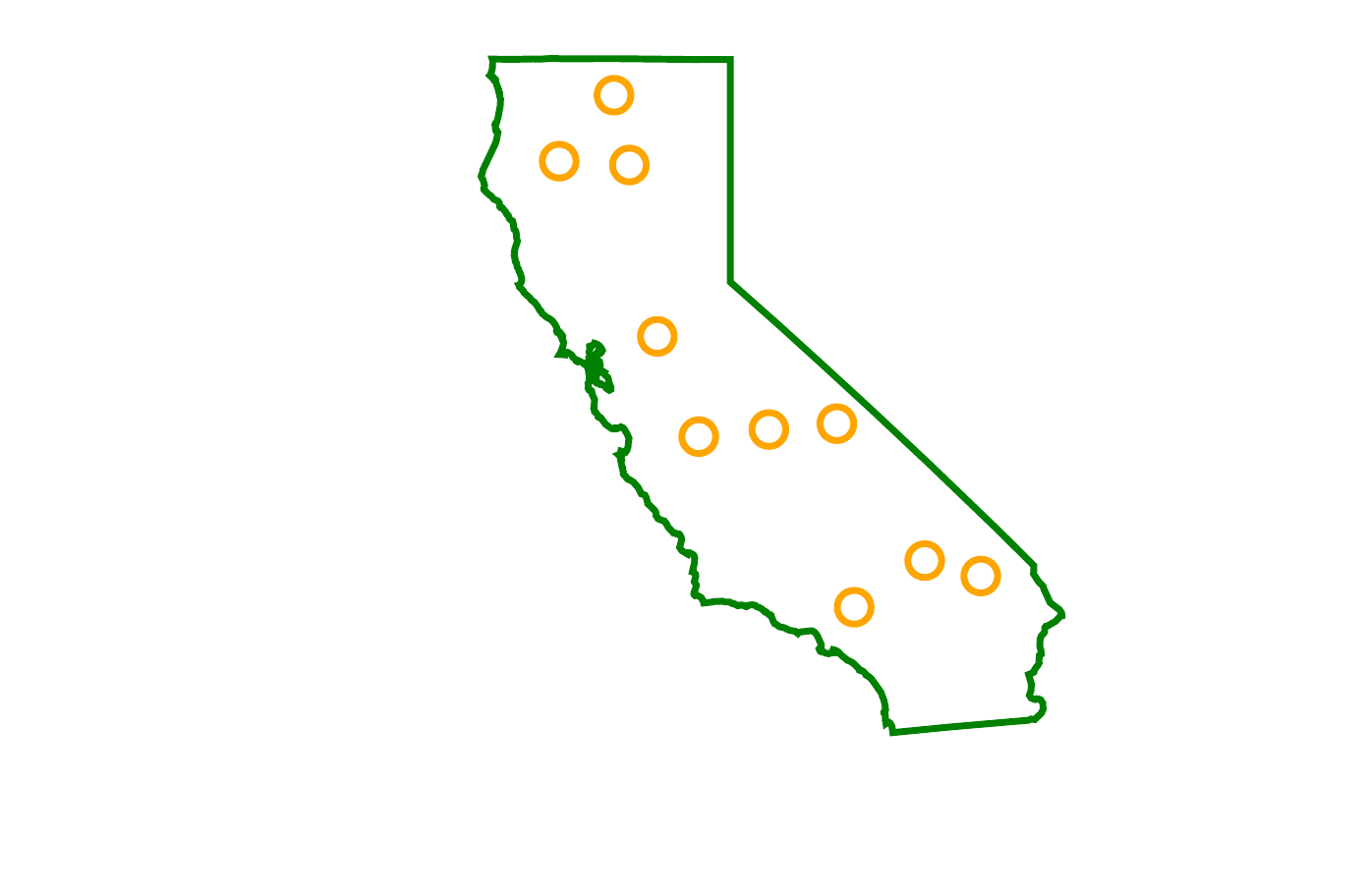}}
    \label{fig:ca_holes}\hfill
  \subfloat[]{%
  \centering
        \includegraphics[width=.36\linewidth]{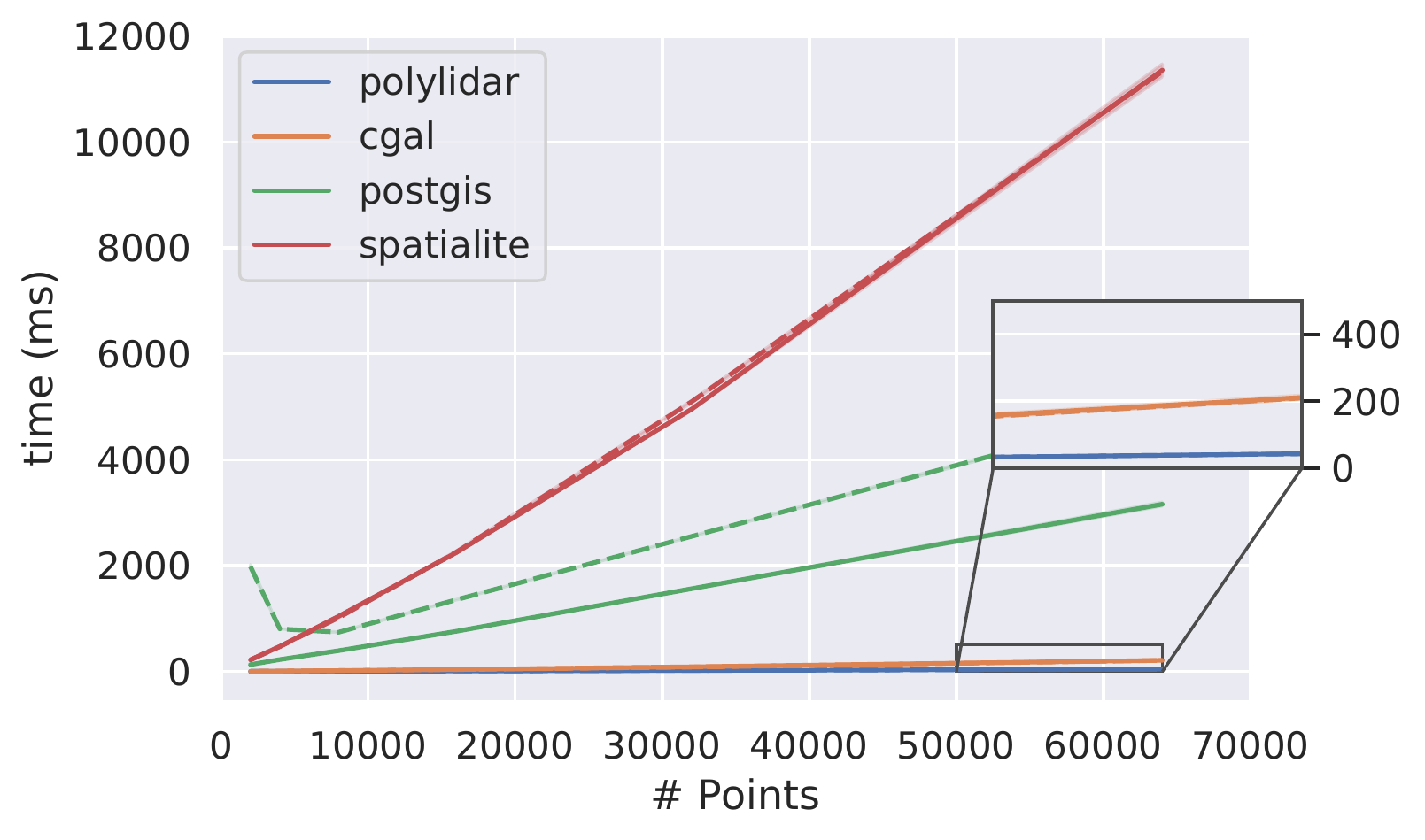}}
    \label{fig:ca_tim}\hfill
  \subfloat[]{%
  \centering
        \includegraphics[width=.33\linewidth]{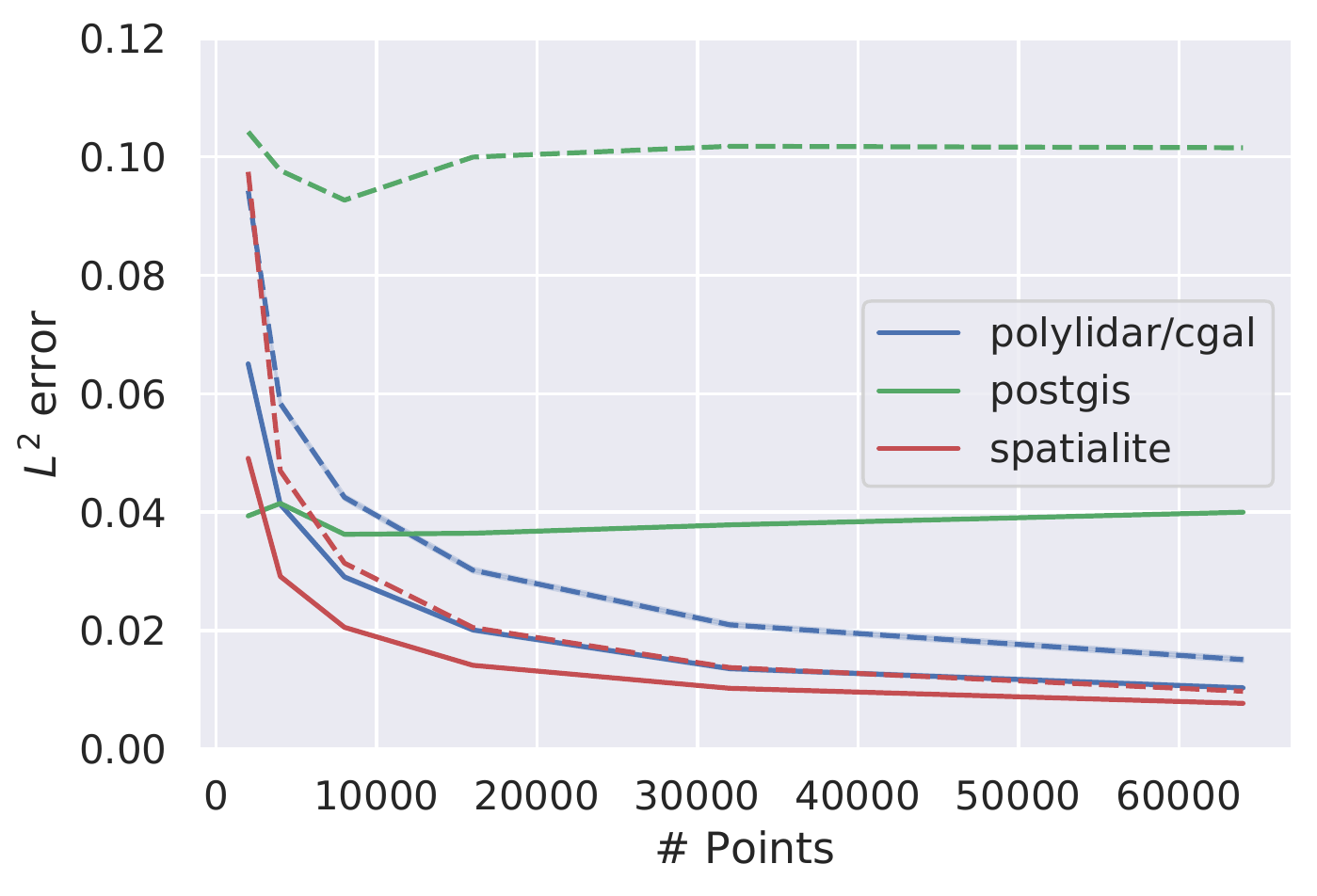}}
    \label{fig:ca_acc}\\
  \subfloat[]{%
      \includegraphics[clip, trim=1.5cm 0cm 1.0cm 0cm, width=0.28\linewidth]{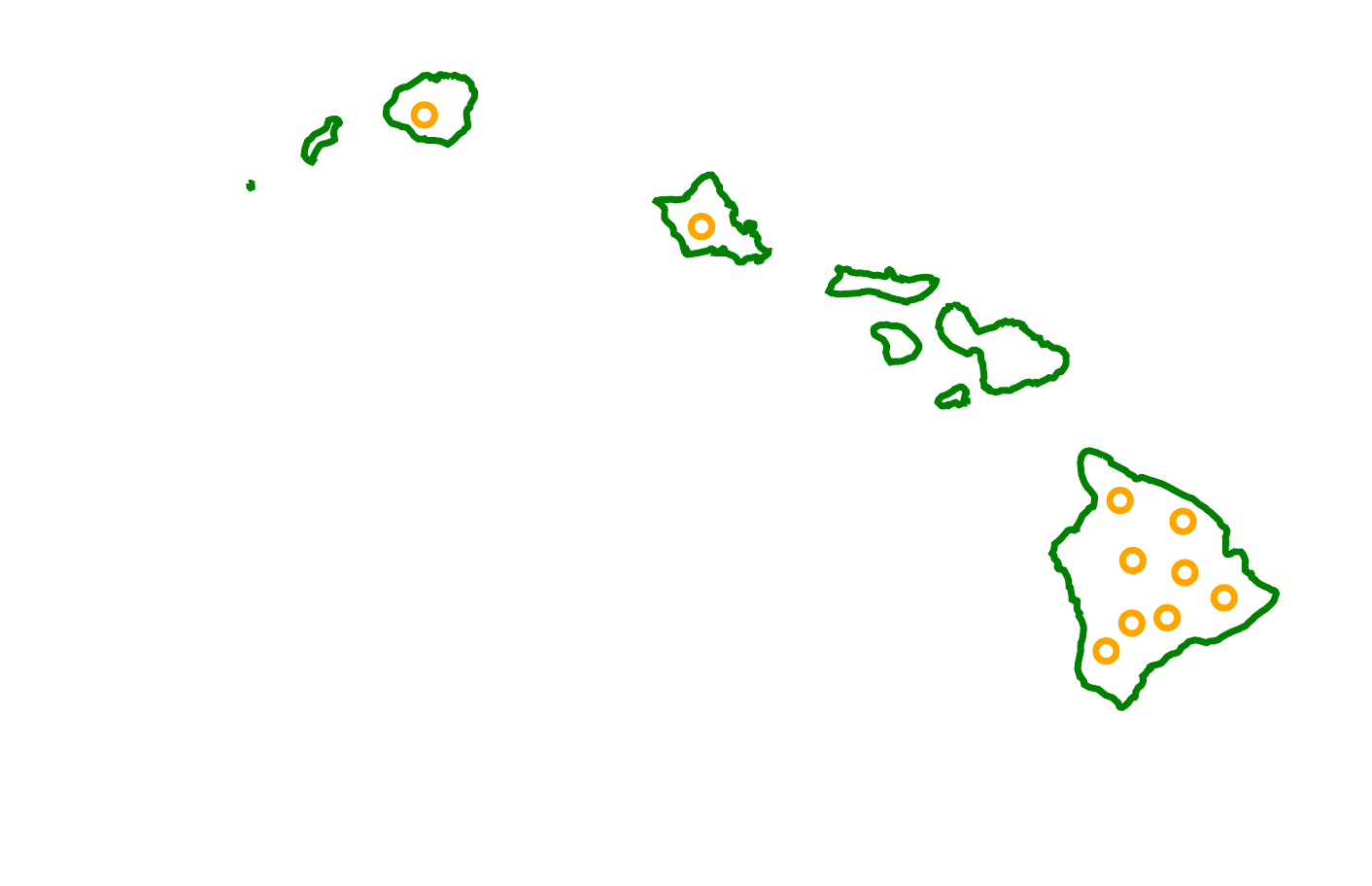}}
    \label{fig:hi_holes}\hfill
  \subfloat[]{%
  \centering
        \includegraphics[width=.36\linewidth]{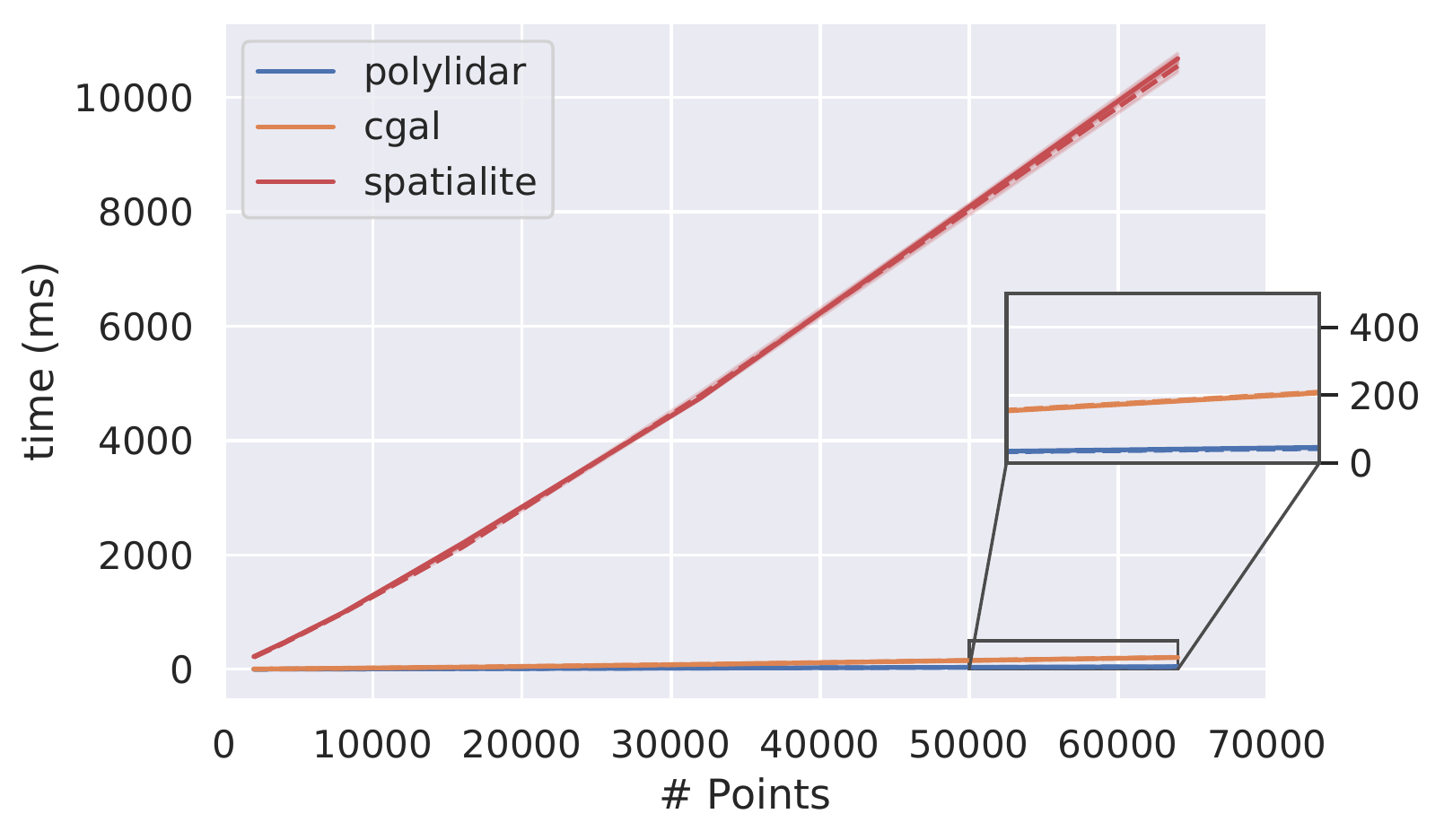}}
    \label{fig:hi_tim}\hfill
  \subfloat[]{%
  \centering
        \includegraphics[width=.33\linewidth]{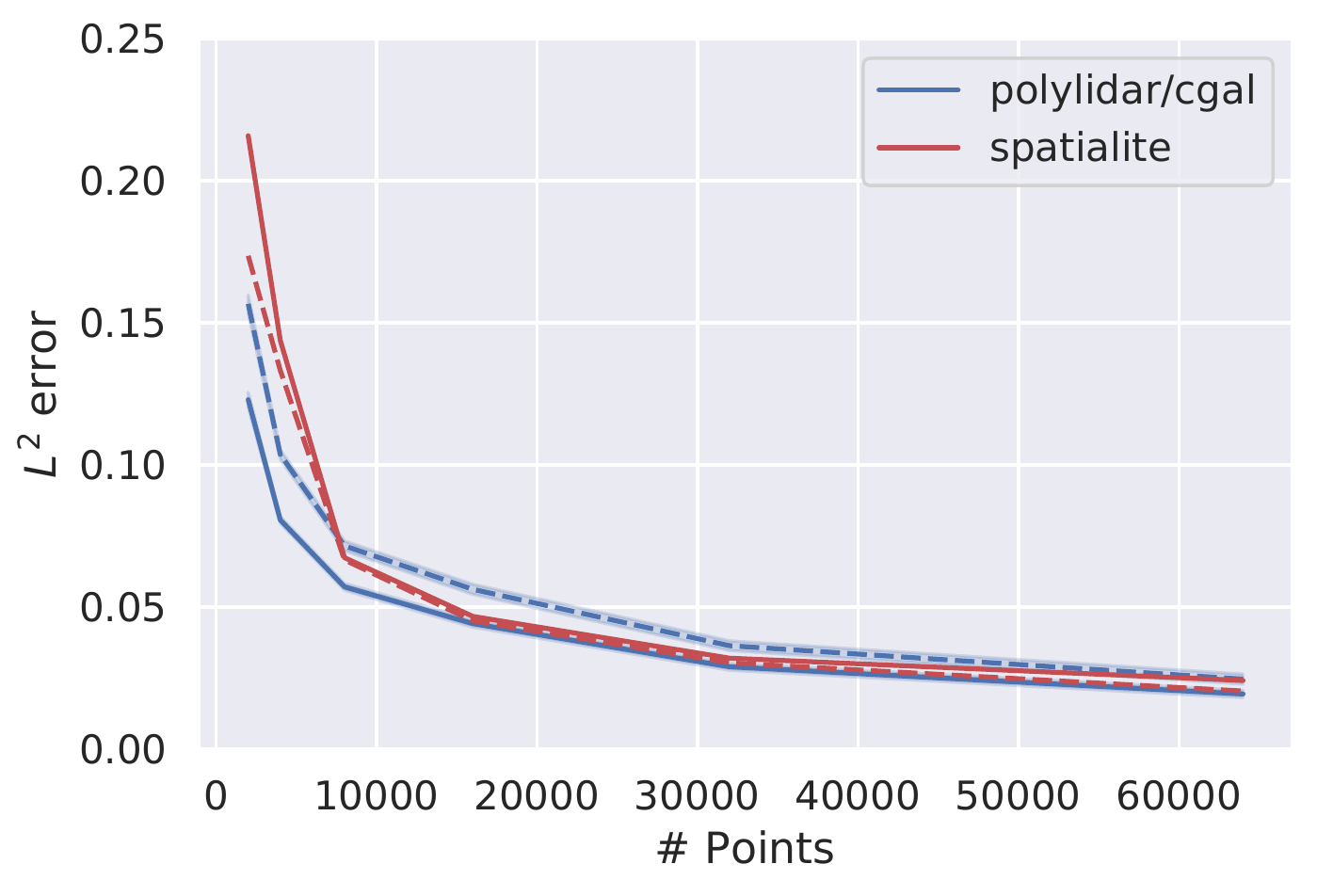}}
    \label{fig:hi_acc}\\
  \caption{Concave hull extraction results. Rows from top to bottom correspond to outlines of California (CA) (a, b, c), and Hawaii (HI) (d, e, f) with random holes inserted. The first column shows ground truth polygons with circular holes in orange. The second column shows execution time as a function of number of 2D points provided. The third column shows shape error as a function of number of 2D points provided. Dashed lines show results where holes were placed inside the polygon outline, while solid lines show results with no holes. PostGIS cannot handle MultiPolygons thus was not tested for HI.}
  \label{fig:compare_algs_all} 
\end{figure*}

\section{Benchmarking Comparisons}\label{sec:results}

This section benchmarks Polylidar against other common concave hull extraction methods which also extract holes; all code is open source\footnote{https://github.com/JeremyBYU/concavehull-evaluation}. Three other implementations are tested: CGAL's Alpha Shape function and the ST\_ConcaveHull function from PostGIS and Spatialite. For uniformity, Polylidar and CGAL are set to use the same $\alpha$ parameter to guarantee exact shape reproduction. Note that CGAL's Alpha Shape returns an unordered set of boundary edges; it does not convert these edges into a valid (multi)polygon. These edges produce the same shape as Polylidar when drawn on a canvas, but lack the desired polygon semantic data structure. PostGIS's concave hull implementation only returns single polygons, so MultiPolygon test cases are not evaluated against it. Both PostGIS and Spatialite are databases which require upload of the point set prior to algorithm execution; benchmark timing does not include data upload time. 

Section \ref{sec:rgbd} provides a benchmark from plane segmented point clouds produced by an RGBD camera. Section \ref{sec:state_shapes} generates synthetic 2D point sets from the state shapes of California (CA) and Hawaii (HI) to explore how the algorithms scale with respect to point size.  Section \ref{sec:alphabet_shapes} shows a similar benchmark but with the English alphabet. All utilize ground truth (multi)polygon shape $GT$ to evaluate shape accuracy. 
Each implementation takes as input a point set and produces a concave shape, $CS$, which is similar to the ground truth polygon.  The $L^2$ error norm, the area of the symmetric difference between $GT$ and $CS$, is computed to enable evaluation of shape error $\frac{area((GT-CS) \cup (CS-GT))}{area(CS)}$. 

Each implementation contains its own parameter(s) modified to minimize $L^2$ error. Shape accuracy is therefore subject to parameter selection. Table \ref{table:params} displays the parameters chosen and used for all test cases (RGBD, CA, HI, Alphabet). Rows with two parameters separated by a semicolon indicate parameters for use with non-hole and hole cases.  Polylidar and CGAL use the same $\alpha$ parameter adjusted on a case by case basis. For each case we calculate point density $p_d$ and compute parameter $\alpha$ as $2p^{-1}_d$.  This gives reasonable but not necessarily optimal results.  Spatialite's concave hull implementation has  parameter $C$ which at its default value ($C=3$) produces excellent results. $C$ is adjusted as needed (for CA, HI) to further reduce error.  PostGIS' \emph{target percent} is set to provide the optimal accuracy based on  percent area reduction required. The most important takeaway when interpreting accuracy is thus trends in accuracy, not small numerical differences.



\begin{table}[h]
\centering
\caption{Parameters for Test Cases}
\label{table:params}
\begin{tabular}{@{}cccccc@{}}
\toprule
Algorithm      & Parameter     & RGBD &       CA          & HI  &  Alph.                  \\ \midrule
CGAL/Polylidar & $\alpha$           & $2p^{-1}_d$  &$2p^{-1}_d$ & $2p^{-1}_d$ & $2p^{-1}_d$   \\ \addlinespace[1mm]
Spatialite     & $C$                & 3.0  &            2.0           & 2.0;1.3 & 3           \\
\addlinespace[1mm]
PostGIS        & $target$ \%  & Varies     & 0.76;0.72   & -         & Varies  \\ \bottomrule
\end{tabular}
\end{table}

\subsection{Plane Segmented Point Clouds from RGBD Images}\label{sec:rgbd}

\begin{figure}[h] 
    \centering
       \includegraphics[width=0.80\linewidth]{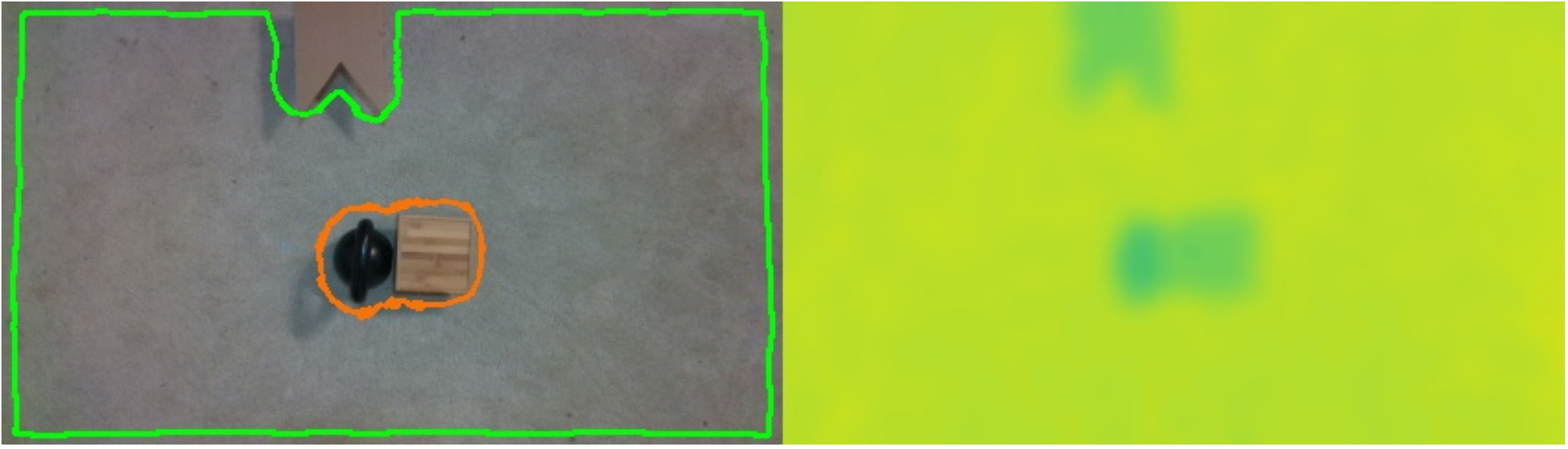}
       \includegraphics[width=0.80\linewidth]{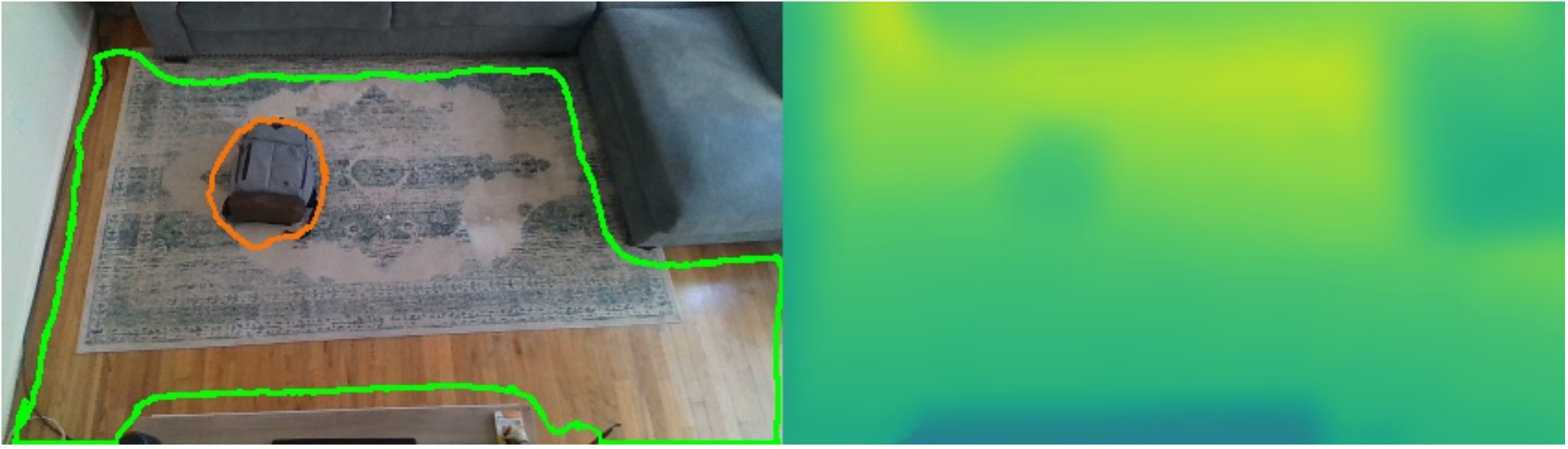}
  \caption{Two example scenes (top/bottom) from RGBD benchmark. A point cloud is generated from depth image (right) and manually segmented to include only the ground floor. The polygonal output of Polylidar is shown in the RGB image (left). Green is the hull, orange represents holes. }
  \label{fig:realsense_benchmark} 
\end{figure}

Point clouds were generated with an Intel RealSense D435i camera at 424X240 resolution from eleven different scenes. Ten scenes were taken with the camera 1.5m above ground level pointing directly downward as shown in the top of Figure \ref{fig:realsense_benchmark}. Floor obstacle positions and orientations were changed in each scene.  The camera was placed higher and angled for the eleventh scene shown in the bottom of Figure \ref{fig:realsense_benchmark}. The floor can be quickly segmented using planar segmentation techniques \cite{feng_fast_2014, pham_geometrically_2016}. However for this experiment the floor was manually segmented, rotated to align with the XY image plane, and subsequently projected. This creates a 2D point set of the floors 3D point cloud. The ground truth polygon of each segmented point cloud was labeled by hand to provide accuracy scores.  The average size of the eleven segmented point clouds is 83,184 points. Table \ref{table:rgbd_results} displays the aggregate execution timings and accuracy results of all eleven points clouds for each algorithm. Polylidar is fastest. Polylidar, CGAL, and Spatialite have similar accuracies. Note that Polylidar and CGAL are configured to produce the same shape and therefore have the same $L^2$ error values.


\begin{table}[!ht]
\centering
\caption{RGBD Plane Segmented Point Clouds}
\label{table:rgbd_results}
\begin{tabular}{lcccccc}
\toprule
{} & \multicolumn{3}{c}{$L^2$ error \%} & \multicolumn{3}{c}{Time (ms)} \\
{Algorithm} &    mean & std &  max &      mean &     std &     max \\
\midrule
Polylidar  &           2.2   &   1.5 &      6.4 & 47.9 & 4.3 & 50.9  \\
CGAL       &           2.2 &    1.5 &      6.4 & 248.3 & 25.0 & 267.7        \\
PostGIS    &           7.5 & 1.6 & 9.9   & 2734.7 & 249.3 & 2939.9     \\
Spatialite &           2.2 & 1.5 & 6.3 &  13333.0 & 2486.6  &   16386.5 \\
\bottomrule
\end{tabular}
\end{table}



\subsection{State Shapes}\label{sec:state_shapes}

Figure \ref{fig:compare_algs_all} shows CA and HI test case geometries (first column), execution times (second column), and error results (third column). Each state shape is processed with and without random holes
; dashed lines indicate results where holes are included in the ground truth polygon. Point sets are randomly sampled from the state shapes. Each test was run 10 times with input point set sizes ranging from (2, 4, 8, 16, 32, 64) thousand points with mean timing and error plotted. Confidence intervals are provided for execution timing, however they are almost imperceptible because the variance is low at this scale. Polylidar and CGAL are significantly faster than the other methods, with Spatialite having the slowest implementation. An inset (zoomed) box that focuses solely on CGAL and Polylidar is shown in the second column, showing that on average Polylidar is $\sim4$ times faster than CGAL. 
The presence of holes affected each method differently: decreased time in Spatialite (fewer triangles to union), increased time for PostGIS (a decrease in \emph{target percent} increases run-time). No significant changes were noted for CGAL and Polylidar.  

\begin{figure}[!ht] 
    \centering
       \includegraphics[clip, trim=0.5cm 1.2cm 0.0cm 0.0cm, width=0.60\linewidth]{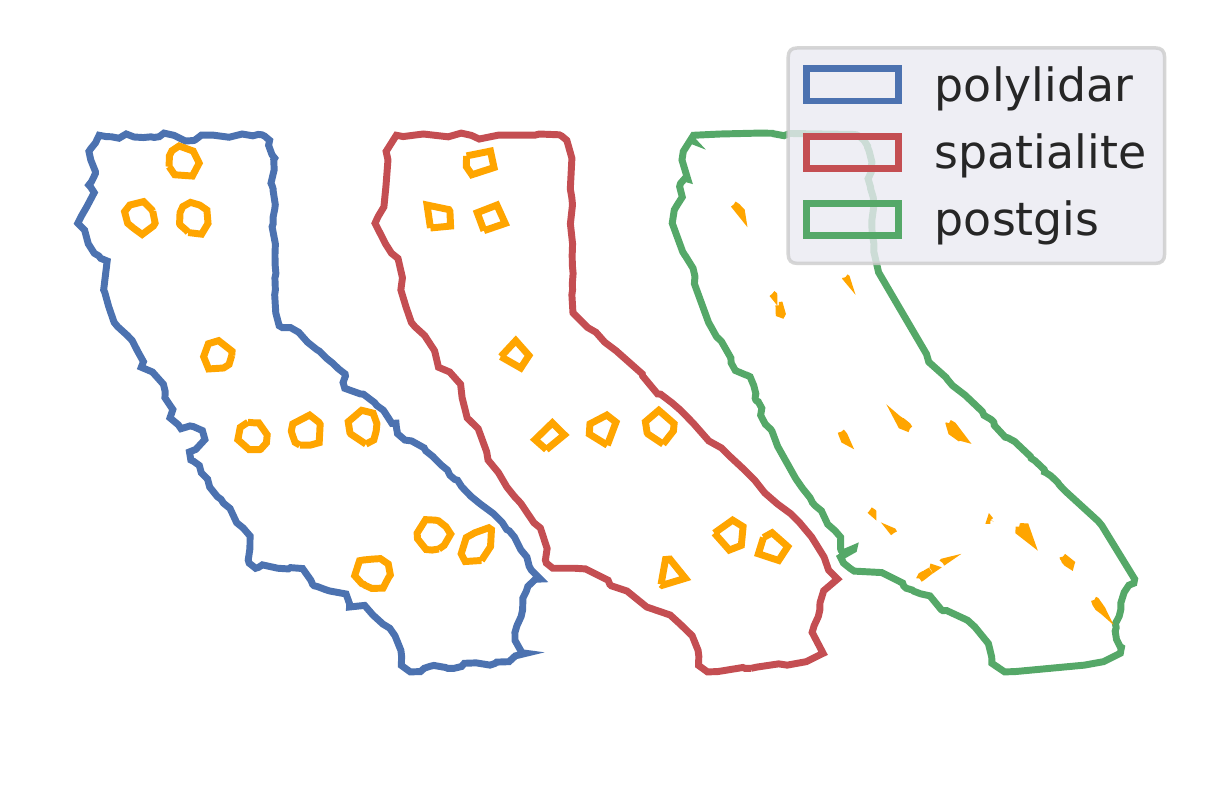}
  \caption{Concave polygon output from Polylidar/CGAL (left), Spatialite (center), and PostGIS (right).  Input to each algorithm was a 4000 point set sampled from the California (CA) polygon with holes per Figure \ref{fig:compare_algs_all}a.}
  \label{fig:ca_output} 
\end{figure}

Spatialite produced shapes with the least error, followed by Polylidar/CGAL and then PostGIS.  Spatialite has the lowest error because it incorporates triangle edge length statistics into its triangle filtering which better handles random sampling. In contrast, Polylidar/CGAL offer comparable accuracies with RGBD data due to the more uniform point distribution in top-down RGBD imagery. PostGIS error increased markedly with holes since it did not accurately reproduce them. 
Figure \ref{fig:ca_output} shows a visual comparison of CA concave polygon outputs for each algorithm. 

\subsection{Alphabet Shapes}\label{sec:alphabet_shapes}

Polygons from 26 capital letters of the English alphabet were generated and 2000 points randomly sampled inside.
The ``A'' in Figure \ref{fig:convex_concave}b shows an example capital letter with the output of Polylidar's concave hull. Table \ref{table:alphabet_tests} provides aggregate statistics of all 26 test cases. Polylidar continues to lead in speed. Spatialite leads in accuracy by a marginal amount. The alphabet shapes are significantly more concave than previous benhchmarks. Documentation of PostGIS indicates that the run time grows quadratically as concavity increases leading to the high execution times observed \cite{postgis}.


 
\begin{table}[!ht]
\centering
\caption{Alphabet Letter Results, 26 Shapes}
\label{table:alphabet_tests}
\begin{tabular}{lcccccc}
\toprule
{} & \multicolumn{3}{c}{$L^2$ error \%} & \multicolumn{3}{c}{Time (ms)} \\
{Algorithm} &    mean & std &  max &      mean &     std &     max \\
\midrule
Polylidar  &           12.8 & 1.8 & 16.8 &       1.2 &    0.3 &     2.4 \\
CGAL       &           12.8 & 1.8 & 16.8 &       5.4 &    0.9 &     7.2 \\
PostGIS    &           36.5 & 9.9 & 53.7 &   13091.8 & 7500.6 & 28451.0 \\
Spatialite &           11.2 & 4.5 & 22.1 &     230.2 &    6.3 &   242.9 \\
\bottomrule
\end{tabular}
\end{table}

\section{Random Polygon Tests}\label{sec:random_polygons_test}
\begin{figure}[!ht] 
    \centering
  \subfloat[]{%
       \includegraphics[width=0.38\linewidth]{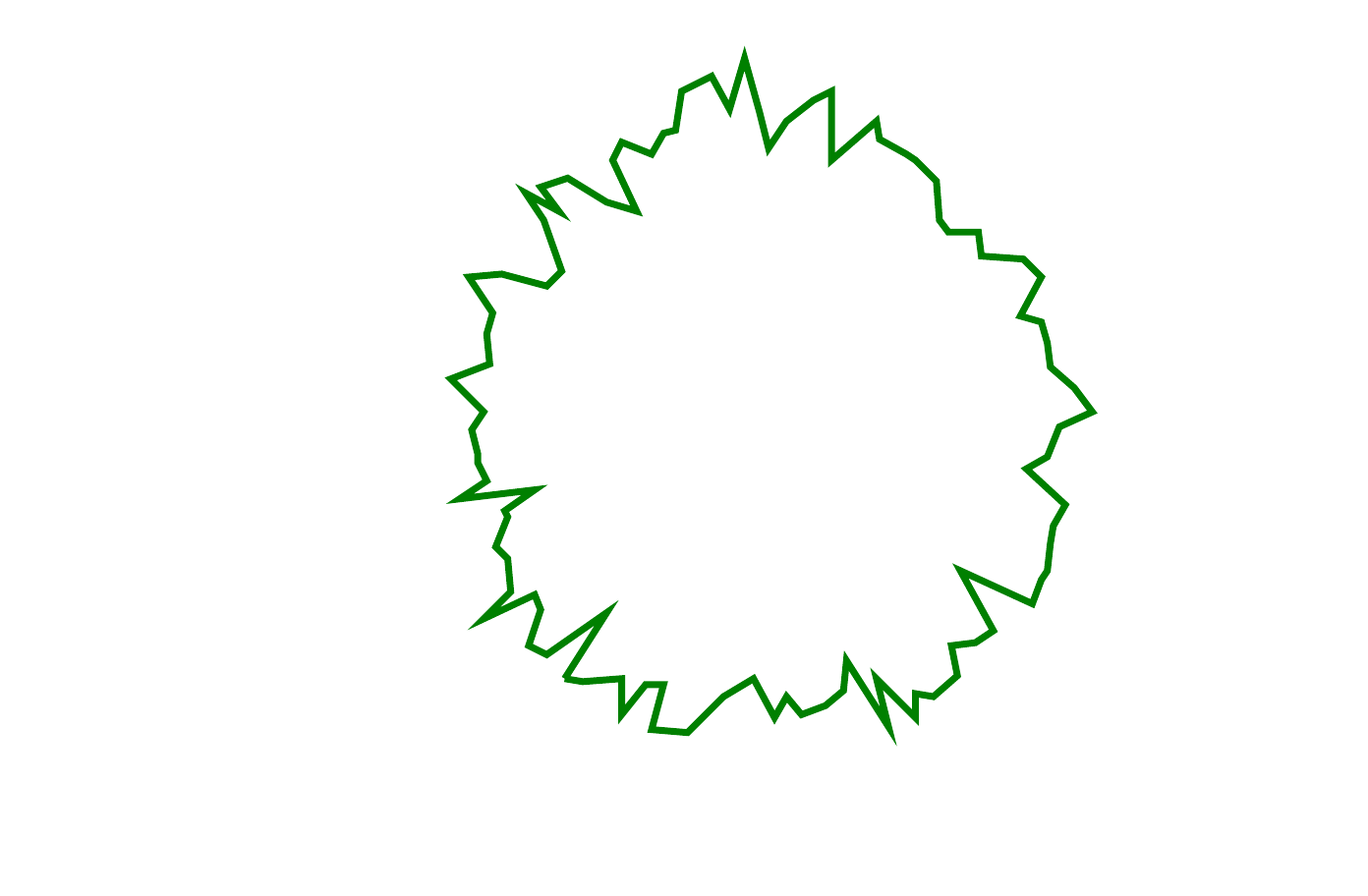}}
    \label{fig:hi_convexity}\hfill
  \subfloat[]{%
        \includegraphics[width=.38\linewidth]{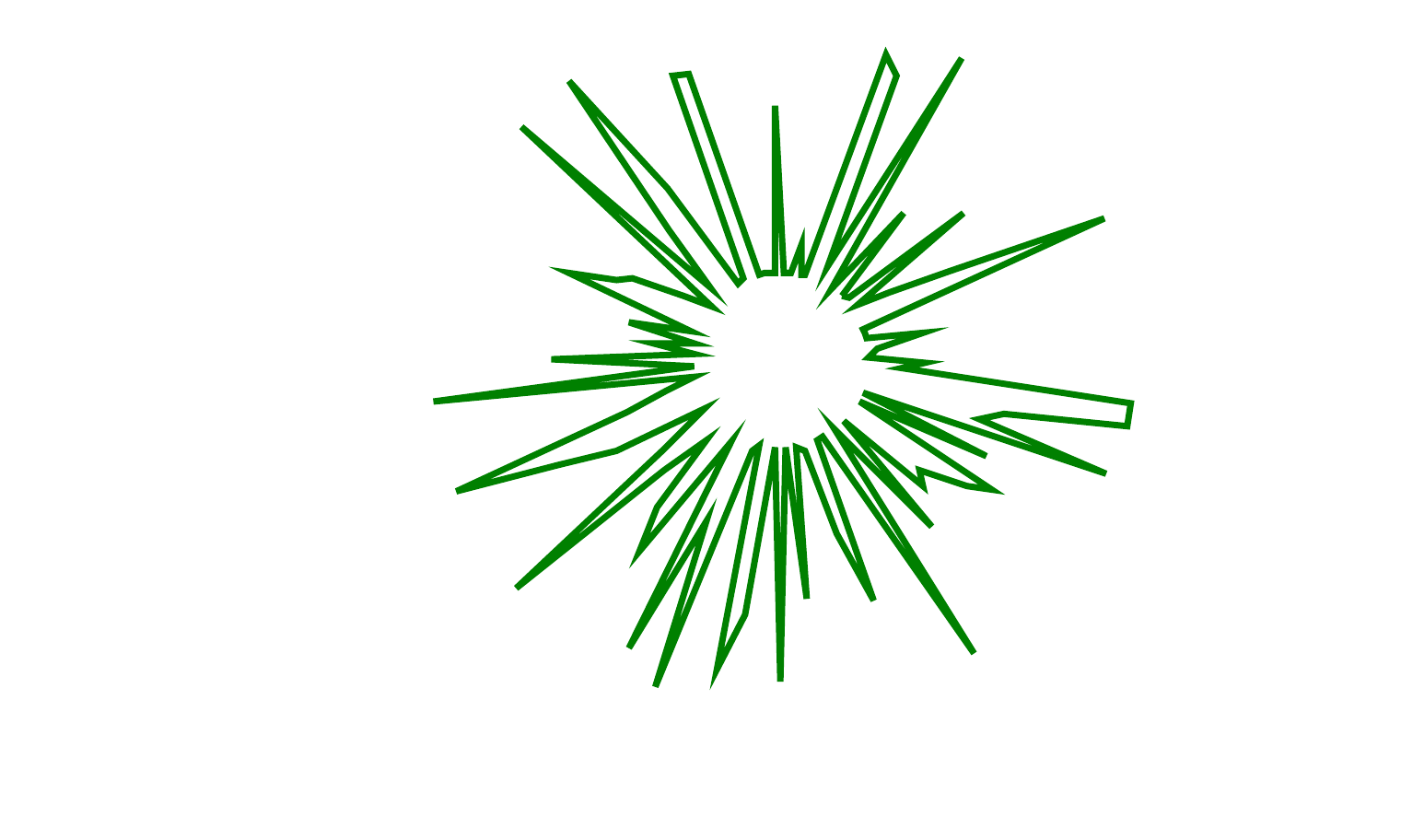}}
    \label{fig:low_convexity}\\
  \caption{(a) Example of a high convexity polygon; CV = 86.1\% (b)  and a low convexity polygon; CV = 26.2\%}
  \label{fig:convexity} 
\end{figure}

More than 19,600 polygons were randomly generated to test Polylidar. Half the test cases had random holes.  Polygon complexity is characterized by convexity metric $$CV = \frac{Area(P)}{Area(CH(P))}$$
\noindent where $P$ is the polygon and $CH()$ is the convex hull function. A convexity of 1 indicates the sample polygon is its convex hull.
8,000 points were randomly sampled for each polygon and input to Polylidar with the $\alpha$ parameter from Table
\ref{table:params}.  Execution time and accuracy are summarized in Table \ref{table:random_tests}. The table is partitioned into high, medium, and low ground truth polygon convexity defined by $CV \geq 0.75$, $0.75 < CV \geq 0.55$, and $CV < 0.55$ respectively. 
Every polygon produced by Polylidar was confirmed valid independently by the GEOS spatial library.
As polygon convexity ($CV$) decreases Polylidar shape estimation accuracy also decreases. Polygons in our ``low'' convexity class have extremely non-convex shapes, the lowest with $CV=0.26$ per Figure \ref{fig:convexity}.

\begin{table}[ht!]
\centering
\caption{Random Tests; $CV$ = Convexity Metric}
\label{table:random_tests}
\begin{tabular}{@{}lcccccc@{}}
\toprule
    &    \multicolumn{3}{c}{$L^2$ error \%} & \multicolumn{3}{c}{Time (ms)} \\
$CV$  &    mean & std &  max &      mean & std &  max \\
\midrule

hi    &     4.4 & 0.5 &  6.1 &       4.6 & 0.1 & 5.0 \\
mid   &     8.0 & 1.1 & 13.0 &       4.6 & 0.1 & 8.1 \\
low   &    15.5 & 3.0 & 25.0 &       4.7 & 0.2 & 9.9 \\
\bottomrule
\end{tabular}
\end{table}

\section{Discussion}\label{sec:discussion}
The benchmarks above indicate that Polylidar is the faster concave (multi)polygon extraction algorithm with the possibility of holes. This section discusses why Polylidar was faster in comparison to others. We specifically analyze the execution time of the major steps in Polylidar in comparison to other triangulation-based methods, namely CGAL and Spatialite. The three major steps are:

\begin{enumerate}
    \item Triangulation - The point set is triangulated creating a mesh of faces, edges, and vertices.
    \item Shape Extraction - Mesh simplices are removed based upon the $\alpha$ parameter or edge length. Remaining triangles, edges, and vertices represent the ``shape".
    \item Polygon Extraction - The ``shape" is converted to a (multi)polygon with the possibility of holes.
\end{enumerate}

\textbf{Triangulation} All perform Delaunay triangulation using robust geometric predicates but use different libraries to do so. Polylidar uses Delaunator, CGAL uses its own 2D triangulation, and Spatialite uses GEOS. 

\textbf{Shape Extraction} Polylidar and Spatialite are most similar, focusing only on filtering triangles in the mesh. However Polylidar goes further with region growing (Section \ref{sec:mesh_extraction}) that isolates disconnected regions in the mesh. For memory efficiency and speed we represent the filtered triangle set $\mathcal{T}_f$ as a bit array with 1/0 indicating in/out of set. This allows rapid triangle filtering and region growing which was previously profiled to be slower when using hashmaps. On the other hand CGAL first creates ``interval hashmaps'' for its simplices, including triangles, edges, and vertices.  These hashmaps store data detailing at what $\alpha$-interval a specific simplex would be in the $\alpha$-complex. These ordered hashmaps give the ability to more quickly compute a \emph{family} of $\alpha$-shapes from a point set. These data structures are implemented as C++ multimaps with $\mathcal{O}(\log{}n)$ for insertion/look-up in comparison to unordered maps having $\mathcal{O}(1)$. This design choice leads to shape extraction having an $\mathcal{O}(n \log{}n)$ complexity for CGAL. By creating hash maps for edges and vertices CGAL can also return the \emph{singular} points and edges which are isolated and not attached to any triangle in the $\alpha$-complex (e.g., a single point far removed from all others). Polylidar need not do this because singular points and edges cannot be polygons thus are not required steps in shape extraction.

\textbf{Polygon Extraction} Polylidar independently converts each region into a polygon. Algorithm 2 quickly identifies all border edges and uses efficient unordered contiguous memory hashmaps to store this information in $\mathcal{HE}$ and $PtE$. The essence of Algorithms 3 and 4 are entirely border-edge based leading to a significant speed up compared to triangle based methods (i.e., perimeter vs. area).  Spatialite uses GEOS to take the union of all unfiltered triangles to generate a valid multipolygon. CGAL's Alpha Shape produces an unordered list of the boundary edges of the $\alpha$-shape. However CGAL does not provide any explicit function to convert this list to a valid (multi)polygon. 


\begin{table}[h]
\centering
\caption{Algorithm Timings - Mean of 30 runs in milliseconds}
\label{table:disc_subtimings}
\begin{tabular}{@{}ccccc@{}}
\toprule
\multirow{2}{*}{Algorithm} & \multirow{2}{*}{triangulation} & \multirow{2}{*}{\begin{tabular}[c]{@{}c@{}}shape\\ extraction\end{tabular}} & \multirow{2}{*}{\begin{tabular}[c]{@{}c@{}}polygon\\ extraction\end{tabular}} & \multirow{2}{*}{total} \\
                           &                                &                                                                             &                                                                               &                        \\ \midrule
Polylidar                  & 36.0                           & 4.4                                                                         & 1.0                                                                           & 41.4                   \\
CGAL                       & 44.5                           & 154.0                                                                       & --                                                                            & 198.5                  \\
Spatialite                 & 234.2                          & 135.3                                                                       & 10788.7                                                                       & 11158.1                \\ \bottomrule
\end{tabular}
\end{table}

Table \ref{table:disc_subtimings} summarizes mean execution timings for each of the main steps for Polylidar, CGAL, and Spatialite. The 64,000 point set in the shape of California (with holes) is used, with each algorithm executed 30 times with the mean presented.  Relative execution times with other point sets are similar.  Delaunator in Polylidar triangulated this specific point set fastest with CGAL a close second. Polylidar achieves a more significant speed-up in shape extraction for which Polylidar is 35 and 32 times faster than CGAL and Spatialite, respectively. Also, Polylidar's polygon extraction is about four orders of magnitude faster than Spatialite whereas CGAL does not extract polygons. CGAL instead offers a general purpose $\alpha$-shape construction routine to compute a family of shapes from different $\alpha$-values. 

\section{Conclusion}\label{sec:conclusion}

This paper has introduced Polylidar, an efficient 2D concave hull extraction algorithm which produces (multi)polygon output with holes. Comparison benchmarks of numerous test sets, similarly done in \cite{Duckham2008}, show Polylidar is faster than competing approaches with comparable or better accuracy. Additionally we perform random polygon tests that confirm every polygon produced by Polylidar is valid.  In future work we will remove Polylidar's reliance on Delaunay triangulation when used with organized point clouds (e.g., RGBD sensors) similar to \cite{lee_fast_2013}. We will explore parallelism to further increase speed and will extend Polylidar to operate directly on 3D point cloud data.




\addtolength{\textheight}{-1cm}   







\bibliographystyle{unsrt}
\bibliography{reference}

\end{document}